# Vector Bessel beams: general classification and scattering simulations

Stefania A. Glukhova[1,2] and Maxim A. Yurkin[1,2,*]

[1] *Voevodsky Institute of Chemical Kinetics and Combustion SB RAS, Institutskaya Str. 3, 630090, Novosibirsk, Russia*

[2] *Novosibirsk State University, Pirogova Str. 2, 630090, Novosibirsk, Russia*

*Corresponding author:* yurkin@gmail.com

## Abstract

Apart from a lot of fundamental interest, vector Bessel beams are widely used in optical manipulation, material processing, and imaging. However, the existing description of such beams remains fragmentary, especially when their scattering by small particles is considered. We propose a new general classification of all existing vortex Bessel beam types in an isotropic medium based on the superposition of transverse Hertz vector potentials. This theoretical framework contains duality and coordinate rotations as elementary matrix operations and naturally describes all relations between various beam types. This leads to various bases for Bessel beams and uncovers the novel beam type with circularly symmetric energy density. We also discuss quadratic functionals of the fields (such as the energy density and Poynting vector) and derive orthogonality relations between various beam types. Altogether, it provides a comprehensive reference of all properties of Bessel beams that may be relevant for applications. Next we generalize the formalism of the Mueller scattering matrices to arbitrary Bessel beams accounting for their vorticity. Finally, we implement these beams in the ADDA code – an open-source parallel implementation of the discrete dipole approximation. This enables easy and efficient simulation of Bessel-beams scattering by particles with arbitrary shape and internal structure.

TOC graphics:

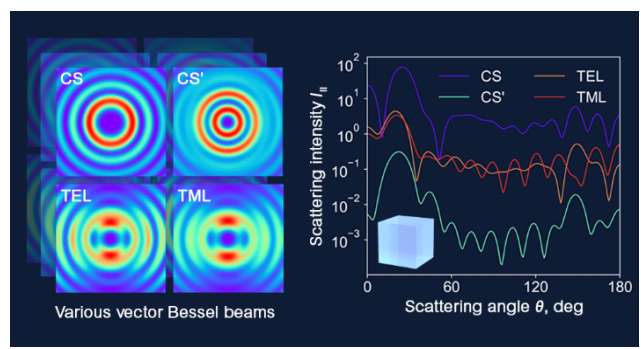

# 1. Introduction

Electromagnetic Bessel beams take prominent positions among other shaped beams and find applications in a vast range of advanced photonics technologies. One of the most remarkable features of the Bessel beam is its lack of diffraction, i.e., the ability to propagate maintaining the profile near the beam axis. Similar to the plane wave, such ideal Bessel beams are not square-integrable (have infinite power flux) and, thus, cannot be produced experimentally. However, using finite apertures allows one to obtain a truncated Bessel beam possessing an "extended focus".[1] Such quasi-Bessel beams can propagate over long distances without significant divergence. A lot of efforts have been devoted to the generation of higher-order and/or vortex Bessel beams using axicons,[2] holograms,[3] or meta-surfaces.[4] Other approaches include radial slot arrays,[5,6] near-field plates,[7] leaky radial waveguides,[8,9] and dielectric planar lenses.[10] The resulting Bessel beams are actively used in such fields as optical manipulation,[11,12] material processing,[13,14] and imaging.[15]

The theoretical description of vector Bessel beams has received much attention over the previous two decades. Several proposed classifications[16–18] were successful in describing some types in a common framework, such as Davis beams derived using the Hertz vector potentials and aplanatic Bessel beams obtained with the angular spectrum decomposition or representation[19]. Despite the conflicting naming of Bessel beam types in the literature, the following groups exist: beams with circularly symmetric energy density (CS type), with transverse electric and magnetic fields – TE and TM types, respectively, and with linear polarizations of electric and magnetic fields – LE and LM beams, respectively. There also exist Bessel beams of fractional order[20,21], but we don't consider it in this paper. Importantly, the previous theoretical descriptions are mostly fragmentary and focus on specific applications. More general classifications[18,21,22] are cumbersome, hiding the relations between various Bessel beam types behind the long formulas. Moreover, the discussion of orthogonality of various beam types or quadratic functionals of the fields, in general, is grossly incomplete. Finally, all of the existing descriptions lack the discussion of rotation transformations of Bessel beams, which is required for the generalization of the scattering-matrix formalism (Mueller calculus[23]) that is commonly used in the case of plane-wave scattering.

Such generalization is desirable, since most of the Bessel beam applications are related to the scattering by (or other linear interaction with) small particles.[24] Another existing limitation is that such scattering has been theoretically considered mostly by particles with spherical symmetry using the generalized Lorenz-Mie theory (GLMT)[25]. The scattering of a zero-order Bessel beam has also been studied for the cases of arbitrary spheroids,[19,26,27] coated spheres,[28] cylinders,[29] large nonspherical homogenous particles,[30] and clusters of spheres.[31–33] By contrast, high-order Bessel beam scattering has been examined only for spheres and spheroids, potentially multilayered and chiral.[34] Importantly, we are not aware of any simulations of Bessel beams for inhomogeneous particles of irregular shape. Moreover, there is a gap between the existing general classification of Bessel beams and scattering simulations for nontrivial particle shapes, i.e., there are no combinations of the latter with complicated beams.

One of the versatile light-scattering simulation methods, applicable to particles with arbitrary shape and internal structure, is the discrete dipole approximation (DDA).[35] Its popularity is based on the conceptual simplicity and availability of well-tested open-source codes, such as DDSCAT[36] and ADDA.[37] In principle, the DDA and corresponding computer codes are applicable to arbitrary incident fields, if it is known at all dipoles (volume discretization elements).[38] However, practical simulations for any beam type are much more accessible to practitioners if these beams are built into the code. Unfortunately, that is currently not the case for any DDA code.

The goal of this paper is two-fold. First, we aim to close the existing knowledge gap by improving the Bessel-beam classification using the Hertz vector potentials, keeping the focus on the interrelations between different beam types and their polarizations. In particular, Section 2 introduces the reader to the Hertz vector potentials and the description of Bessel beam types known in the literature but rewritten in a new form. In Section 3 we propose a general classification of Bessel beams in any homogeneous isotropic medium based on a 2×2 matrix of coefficients and thoroughly discuss its various properties. The latter range from rotation and duality transformations (Sections 3.2 and 3.3) to the description of quadratic functionals of the fields, such as the energy density and Poynting vector (Section 3.4), and discussion of various bases of Bessel beam including their orthogonality (Section 3.5).

The second goal is to implement all types of Bessel in the open-source ADDA code, keeping it compatible with the Mueller calculus (Section 4.1). The details of this implementation are given in Section 4.2, while in Section 5 we verify our code and provide several simulation examples. Section 6 concludes the paper. Preliminary results of this work have been reported at the conferences.[39,40]

## 2. Existing Bessel beam types

### 2.1. Hertz vector potentials

We consider monochromatic electromagnetic waves (beams) with the time dependence $\exp(-i\omega t)$ in a homogeneous isotropic medium (with absolute permittivity $\varepsilon$, permeability $\mu$, and wave impedance $\eta \stackrel{\text{def}}{=} \sqrt{\mu/\varepsilon}$) without explicit consideration of the sources that produce them. In other words, we assume the sources to be located at infinity. We use SI units in this paper, although this makes some of expressions a bit more complicated than that for commonly used Gaussian units. The Maxwell's equations are then given by:[41]

$$\nabla \times \mathbf{E}(\mathbf{r}) = i\omega\mu\mathbf{H}(\mathbf{r}),$$
$$\nabla \times \mathbf{H}(\mathbf{r}) = -i\omega\varepsilon\mathbf{E}(\mathbf{r}), \tag{1}$$

which implies that both electric and magnetic fields (**E** and **H**) satisfy a homogeneous vector Helmholtz equation:

$$\nabla^2 \mathbf{E}(\mathbf{r}) + k^2 \mathbf{E}(\mathbf{r}) = 0, \tag{2}$$

$$\nabla \times \nabla \times \mathbf{E}(\mathbf{r}) - k^2 \mathbf{E}(\mathbf{r}) = 0, \tag{3}$$

where $k \stackrel{\text{def}}{=} \omega\sqrt{\varepsilon\mu}$ is the wave number, and the two equations are equivalent due $\nabla \cdot \mathbf{E}(\mathbf{r}) = 0$. However, Eq. (3) [but not Eq. (2)] implies $\nabla \cdot \mathbf{E}(\mathbf{r}) = 0$ and is, thus, alone equivalent to Eq. (1). Here and further we present expressions mostly for $\mathbf{E}(\mathbf{r})$, since it uniquely determines $\mathbf{H}(\mathbf{r})$ through Eq. (1). We also further omit the common dependence of fields, potentials, etc. on **r** for brevity. We consider arbitrary $k$, including complex ones. The strongly absorbing host medium implies that the fields decay over a small distance, which is incompatible with the field sources located far from the scatterer. However, a weakly absorbing medium (relatively small imaginary part of $k$) is compatible with such scattering problem. But practical aspects of generating Bessel beams in such host medium is outside the scope of this paper.

To describe various Bessel beams, it is convenient to use the Hertz vector potentials.[16] They are related to the scalar and vector potentials ($\phi$, **A**) that define **E** and **H** through:

$$\mathbf{E} = -\nabla\phi + i\omega\mathbf{A}, \tag{4}$$

$$\mathbf{H} = \frac{1}{\mu}\nabla \times \mathbf{A}. \tag{5}$$

Electric and magnetic Hertz vector potentials $\mathbf{\Pi}_e$ and $\mathbf{\Pi}_m$ are introduced as:[42]

$$\mathbf{A} = -ik^2\omega^{-1}\mathbf{\Pi}_e + \mu\nabla \times \mathbf{\Pi}_m, \tag{6}$$

$$\phi = -\nabla \cdot \mathbf{\Pi}_e, \tag{7}$$

which implies the Lorenz gauge condition for $\phi$ and **A** and that these potentials satisfy scalar and vector Helmholtz equations, respectively [Eq. (2)]. Note, however, that this gauge condition uniquely determines potentials only if they are additionally required to decay sufficiently fast at infinity.[43] In source-free case as considered here, this would imply vanishing of both potentials and fields in the whole space. Equations (4)–(7) imply the following expressions for the fields:

$$\mathbf{E} = \nabla\nabla \cdot \mathbf{\Pi}_e + k^2\mathbf{\Pi}_e + ik\eta\nabla \times \mathbf{\Pi}_m, \tag{8}$$

$$\mathbf{H} = \nabla \times (\nabla \times \mathbf{\Pi}_m - ik\eta^{-1}\mathbf{\Pi}_e), \tag{9}$$

which leaves a lot of freedom in choosing $\mathbf{\Pi}_e$ and $\mathbf{\Pi}_m$ for given fields. The first restriction, which we postulate is that both $\mathbf{\Pi}_e$ and $\mathbf{\Pi}_m$ satisfy vector Helmholtz equation [Eq. (2)].[43] This allows us to rewrite Eq. (8) in a form symmetric to Eq. (9)

$$\mathbf{E} = \nabla \times (\nabla \times \mathbf{\Pi}_e + ik\eta\mathbf{\Pi}_m). \tag{10}$$

Conversely, any $\mathbf{\Pi}_e$ and $\mathbf{\Pi}_m$ – solutions to Eq. (2), being substituted into Eqs. (8), (9), lead to **E** and **H** satisfying Eq. (1).

Let us briefly mention the remaining degrees of freedom, which we do not fix. First, for any function $g$, satisfying the scalar Helmholtz equation

$$\nabla^2 g + k^2 g = 0, \tag{11}$$

$\mathbf{\Pi}_e$ (or $\mathbf{\Pi}_m$) can be incremented by $\nabla g$ without changing the fields, but decrementing $\nabla \cdot \mathbf{\Pi}_e$ (or $\nabla \cdot \mathbf{\Pi}_m$) by $k^2 g$. Such variation of $\mathbf{\Pi}_e$ (but not of $\mathbf{\Pi}_m$) also modifies $\phi$ and **A** without violating the Lorenz gauge. In other words, only the curls of those vector potentials determine the fields. Second, the fields are not affected if $\mathbf{\Pi}_e$ and $\mathbf{\Pi}_m$ are simultaneously incremented by any auxiliary fields $\mathbf{E}'$ and $\mathbf{H}'$, respectively, if these fields satisfy Eq. (1). Combining the above options, either $\mathbf{\Pi}_e$ and $\mathbf{\Pi}_m$ can be set to zero – then the other one will equal to the corresponding (electric or magnetic) field divided by $k^2$.

For Bessel beams propagating along the $z$-axis, both $\mathbf{\Pi}_e$ and $\mathbf{\Pi}_m$ has the following simple functional form in the cylindrical coordinate system[17]

$$f_n(\mathbf{r}) = J_n(k_t \rho) e^{in\varphi} e^{ik_z z}, \tag{12}$$

where $J_n$ is the Bessel function of the first kind ($n$ is the order of the Bessel beam), $k_t \stackrel{\text{def}}{=} k \sin \alpha_0$ and $k_z \stackrel{\text{def}}{=} k \cos \alpha_0$ are the transverse and longitudinal components of the wave vector $k$, respectively, and $\alpha_0$ is the half cone angle (**Figure 1**). More generally, $k_t$ and $k_z$ can be arbitrary complex values with the only constraint $k_t^2 + k_z^2 = k^2$, but most applications consider real angles $\alpha_0$, i.e. the ratios of any two of these wave numbers is real. Note that some researchers[30,32,44] introduce additional factor $i^n$ into the definition of $f_n$ [Eq. (12)], which affects constant factors in all further relations where several orders of $f_n$ (or corresponding potentials or fields) are present. We consider only integer values of $n$ in this paper, since otherwise $f_n$ is discontinuous with respect to $\varphi$. Importantly, $f_n$ is the solution to Eq. (11); thus, any linear superposition of $f_n$ with constant vectors (and potentially different $n$ and $\alpha_0$) satisfies Eq. (2) and qualifies as either $\mathbf{\Pi}_e$ or $\mathbf{\Pi}_m$.

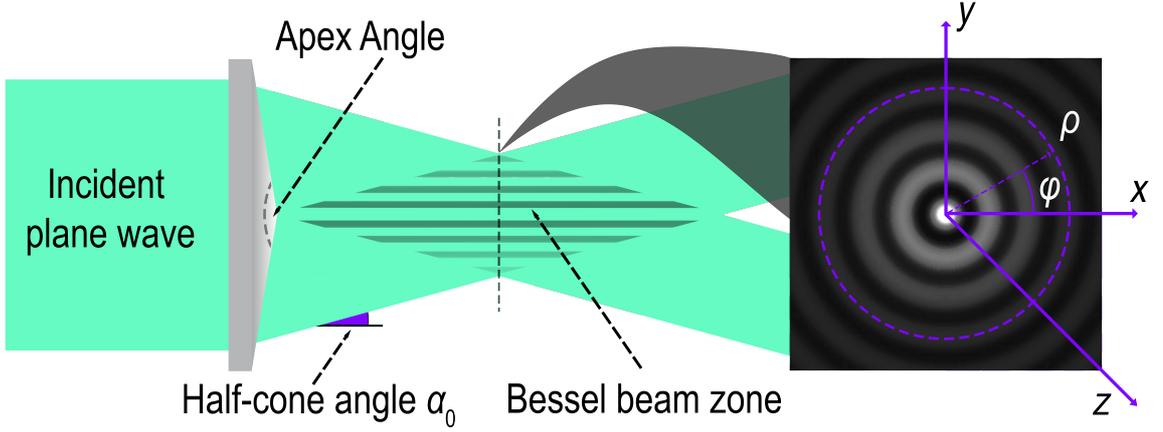

**Figure 1.** Geometry of Bessel beam production using an axicon (conical lens). In the limit $\alpha_0 \to 0$, the beam is a plane wave. Adapted with permission.[17] Copyright 2016, Elsevier.

Before continuing let us introduce a convenient notation $\mathbf{e}_\pm \stackrel{\text{def}}{=} \mathbf{e}_x \pm i\mathbf{e}_y$ (where $\mathbf{e}_x$ and $\mathbf{e}_y$ are unit vectors along the corresponding axes), having the following properties

$$\mathbf{e}_\pm \times \mathbf{e}_z = \pm i\mathbf{e}_\pm, \quad \mathbf{e}_\mp \times \mathbf{e}_\pm = \pm 2i\mathbf{e}_z, \quad \mathbf{e}_\pm \times \mathbf{e}_\pm = \mathbf{e}_\pm \cdot \mathbf{e}_\pm = 0, \quad \mathbf{e}_\pm \cdot \mathbf{e}_\mp = 2. \quad (13)$$

Note that we define the (bilinear) dot product without conjugation of the second argument to be compatible with the notation for divergence, i.e., this is not a proper inner product of two complex vectors. The latter can be obtained with the help of trivial relation $\mathbf{e}_\pm^* = \mathbf{e}_\mp$, where * denotes complex conjugate. Similar circular unit vectors are common, e.g., in description of vector spherical harmonics, e.g.[17], but here we keep them not normalized (by a factor $\sqrt{2}$) to simplify further expressions. Next, elementary calculus with recurrent relations for Bessel functions leads to the following identities:

$$\nabla f_n = ik_z f_n \mathbf{e}_z + \frac{k_t}{2}(f_{n-1}\mathbf{e}_+ - f_{n+1}\mathbf{e}_-), \quad (14)$$

$$\nabla \times (f_n \mathbf{e}_z) = \frac{ik_t}{2}(f_{n-1}\mathbf{e}_+ + f_{n+1}\mathbf{e}_-), \quad \nabla \times (f_n \mathbf{e}_\pm) = \pm k_z f_n \mathbf{e}_\pm - ik_t f_{n\pm 1}\mathbf{e}_z, \quad (15)$$

$$\nabla \cdot (f_n \mathbf{e}_z) = ik_z f_n, \quad \nabla \cdot (f_n \mathbf{e}_\pm) = \mp k_t f_{n\pm 1}, \quad (16)$$

$$\nabla \times \nabla \times (f_n \mathbf{e}_z) = ik_z \nabla f_n + k^2 f_n \mathbf{e}_z = k_t^2 f_n \mathbf{e}_z + i\frac{k_z k_t}{2}(f_{n-1}\mathbf{e}_+ - f_{n+1}\mathbf{e}_-), \quad (17)$$

$$\nabla \times \nabla \times (f_n \mathbf{e}_\pm) = \mp k_t \nabla f_{n\pm 1} + k^2 f_n \mathbf{e}_\pm = \frac{k^2 + k_z^2}{2} f_n \mathbf{e}_\pm \mp ik_z k_t f_{n\pm 1}\mathbf{e}_z + \frac{k_t^2}{2} f_{n\pm 2}\mathbf{e}_\mp. \quad (18)$$

We are not aware of previous explicit discussion of these relations, but they make most of the following expressions and derivations substantially more concise. Note that curls and double curls in Eqs. (15), (17), (18) are divergence-free and act as building blocks for the fields due to Eqs. (8), (9), as illustrated in the next section. However, they are not all linearly independent, if several orders $n$ are considered together (see Section 3). In particular, $\nabla \times (f_n \mathbf{e}_z)$ and $\nabla \times \nabla \times (f_n \mathbf{e}_z)$ are the cylindrical vector wave functions (CVWFs), which are known to be a complete basis set for expansion of electromagnetic field in free space.[45,46]

## 2.2. Davis description of Bessel beams

Varying the direction of the Hertz vector potentials, we obtain different beam types,[16] which are sometimes called Davis formulation of Bessel beams in the literature.[18] The following field expressions can be easily derived using Eqs. (13)–(18).

TE and TM (transverse electric and magnetic) Bessel beams are obtained from $\mathbf{\Pi}_m = \eta^{-1}\Pi_0 f_n \mathbf{e}_z$, $\mathbf{\Pi}_e = 0$ and $\mathbf{\Pi}_e = \Pi_0 f_n \mathbf{e}_z$, $\mathbf{\Pi}_m = 0$, respectively. We denote the corresponding fields $\mathbf{E}_{\mathrm{TE}}$ and $\mathbf{H}_{\mathrm{TM}}$, respectively; they have zero $z$-components. The accompanying fields $\mathbf{H}_{\mathrm{TE}}$ and $\mathbf{E}_{\mathrm{TM}}$ generally have no zero components. The amplitudes (scaling factors) for $\mathbf{\Pi}_e$ are related to that for the electric field, as $\Pi_0 = E_0/k^2$. Here and further we omit the index $n$ for fields and potentials to avoid clutter, unless several orders appear in the same equation. The electric fields of the TE and TM beams are the following:

$$E_{\mathrm{TE},x} = -\frac{E_0 k_t}{2k}(f_{n-1} + f_{n+1}), \qquad E_{\mathrm{TE},y} = \mathrm{i}\frac{E_0 k_t}{2k}(f_{n+1} - f_{n-1}), \qquad E_{\mathrm{TE},z} = 0, \qquad (19)$$

$$E_{\mathrm{TM},x} = \mathrm{i}\frac{E_0 k_t k_z}{2k^2}(f_{n-1} - f_{n+1}), \qquad E_{\mathrm{TM},y} = -\frac{E_0 k_t k_z}{2k^2}(f_{n-1} + f_{n+1}),$$

$$E_{\mathrm{TM},z} = \frac{E_0 k_t^2}{k^2} f_n. \qquad (20)$$

Magnitude profiles for the electric field and time-averaged Poynting vector (both separate components and total vector magnitude), as well as profiles of time-averaged energy density of the TE and TM beams are presented in Figure 2 and Figure 3. Note that these beams have identical Poynting vectors and energy densities due to the duality transformation (see Section 3.3). Moreover, the rigorous definition and discussion of these quadratic functionals of the fields are deferred to Section 3.4.

Mathematically, these Bessel beams are the most fundamental ones, since they are directly proportional to the CVWFs. They are very convenient in reflection and transmission problems.[47] Moreover, such beams of zero order are connected with the azimuthal and radial beam polarizations.[16]

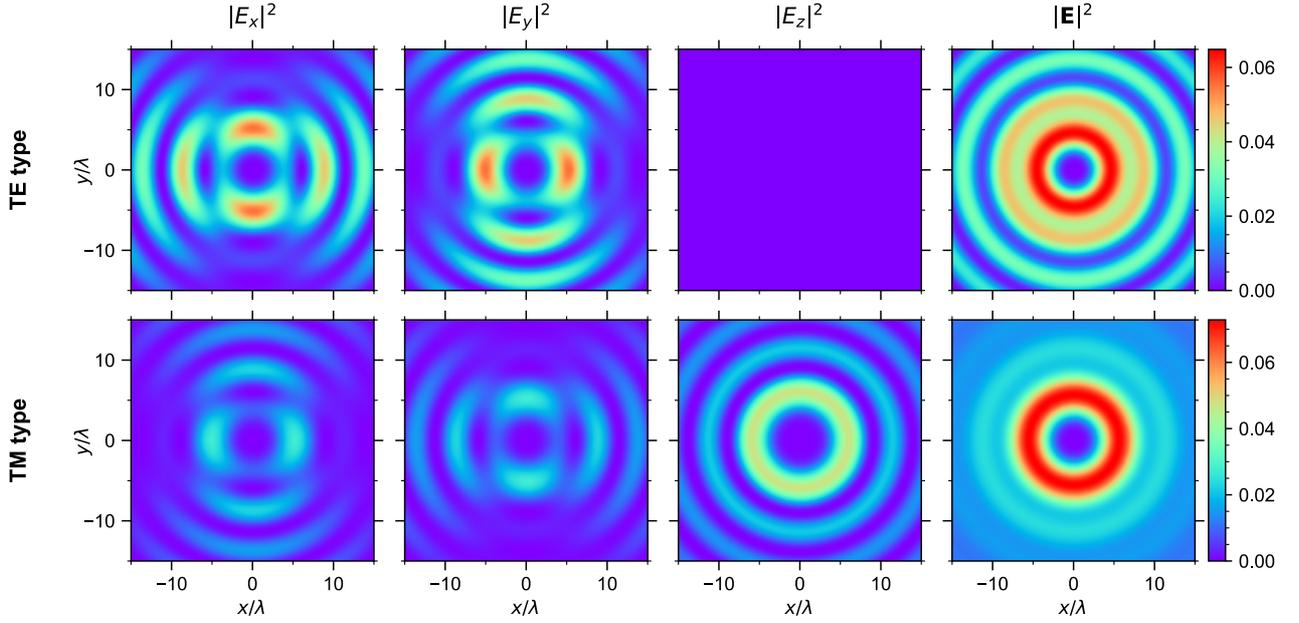

**Figure 2.** Intensity profiles of components of $\mathbf{E}_{TE}$ and $\mathbf{E}_{TM}$ with $n = 2$, $\alpha_0 = 45^0$ in the $xy$-plane. The amplitude $E_0$ and wavenumber $k$ are equal to 1, the $z$-component of $\mathbf{E}_{TE}$ is identically zero [see Eq. (19)].

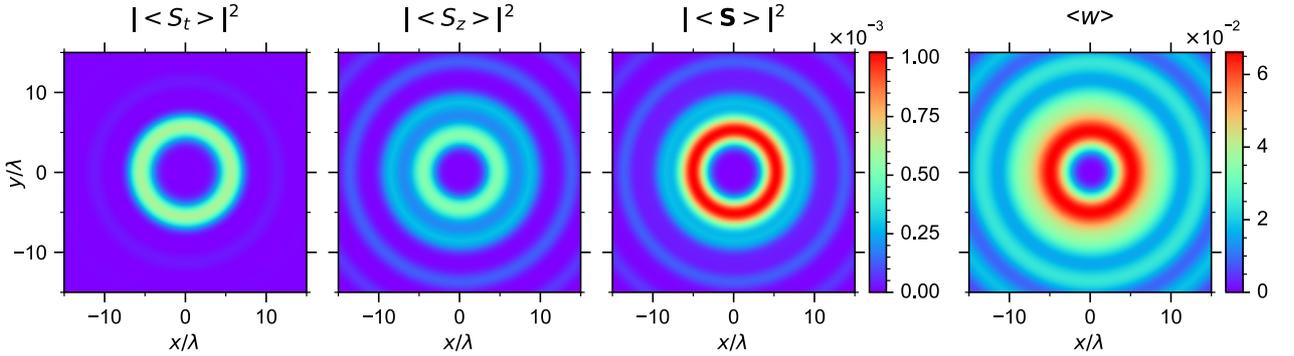

**Figure 3.** Magnitude profiles of the Poynting vector components and energy density for either $\mathbf{E}_{TE}$ or $\mathbf{E}_{TM}$ with the same parameters as in Figure 2.

Bessel beams with linearly polarized electric and magnetic fields (LE and LM, respectively) are obtained from $\mathbf{\Pi}_m = \eta^{-1}\Pi_0 f_n \mathbf{e}_t$, $\mathbf{\Pi}_e = 0$ and $\mathbf{\Pi}_e = \Pi_0 f_n \mathbf{e}_t$, $\mathbf{\Pi}_m = 0$, respectively. Here $\mathbf{e}_t$ is a transverse polarization vector perpendicular to $\mathbf{e}_z$. When it equals $\mathbf{e}_z \times \mathbf{e}_{x,y}$ (i.e., $\mathbf{e}_y$ or $-\mathbf{e}_x$), it leads to so-called $x$- or $y$-linear polarizations of the corresponding fields, respectively, which are denoted as $\mathbf{E}_m^{(x)}$, $\mathbf{E}_m^{(y)}$ and $\mathbf{H}_e^{(x)}$, $\mathbf{H}_e^{(y)}$ for LE and LM fields in the literature.[17] Here subscripts m and e correspond to the type of non-zero Hertz vector potential, and superscripts $(x)$, $(y)$ denoting polarization should not be confused with subscripts $x$, $y$, $z$ denoting vector components. However, we prefer to use subscripts LE and LM in the field expressions to keep uniform notation across all Bessel beam types. Moreover, our definition of $(y)$ polarization ($\mathbf{e}_t = -\mathbf{e}_x$) corresponds to the linear $y$-polarization of a plane wave in the limit of $\alpha_0 = 0°$ and $n = 0$ (see Section 3.2), but has inverse sign to that in Ref. [17]. The resulting electric fields for the $x$-polarized LE and LM Bessel beams are the following:

$$E_{LE,x}^{(x)} = \frac{E_0 k_z}{k} f_n, \qquad E_{LE,y}^{(x)} = 0, \qquad E_{LE,z}^{(x)} = i\frac{E_0 k_t}{2k}(f_{n-1} - f_{n+1}), \tag{21}$$

$$E_{LM,x}^{(x)} = i\frac{E_0 k_t^2}{4k^2}(f_{n-2} - f_{n+2}),$$

$$E_{LM,y}^{(x)} = \frac{E_0}{k^2}\left[\frac{k^2 + k_z^2}{2} f_n - \frac{k_t^2}{4}(f_{n-2} + f_{n+2})\right], \qquad E_{LM,z}^{(x)} = -\frac{E_0 k_t k_z}{2k^2}(f_{n-1} + f_{n+1}). \tag{22}$$

Magnitude profiles for the electric field, Poynting vector, and energy density of LE and LM types are presented in Figure 4 and Figure 5. Note that these fields always have non-zero longitudinal components, however the electric (for LE type) or magnetic (for LM type) field has zero component along $\mathbf{e}_t$. The accompanying magnetic $\mathbf{H}_{LE}^{(x)}$, $\mathbf{H}_{LE}^{(y)}$ and electric fields $\mathbf{E}_{LM}^{(x)}$, $\mathbf{E}_{LM}^{(y)}$ have no zero components at all.

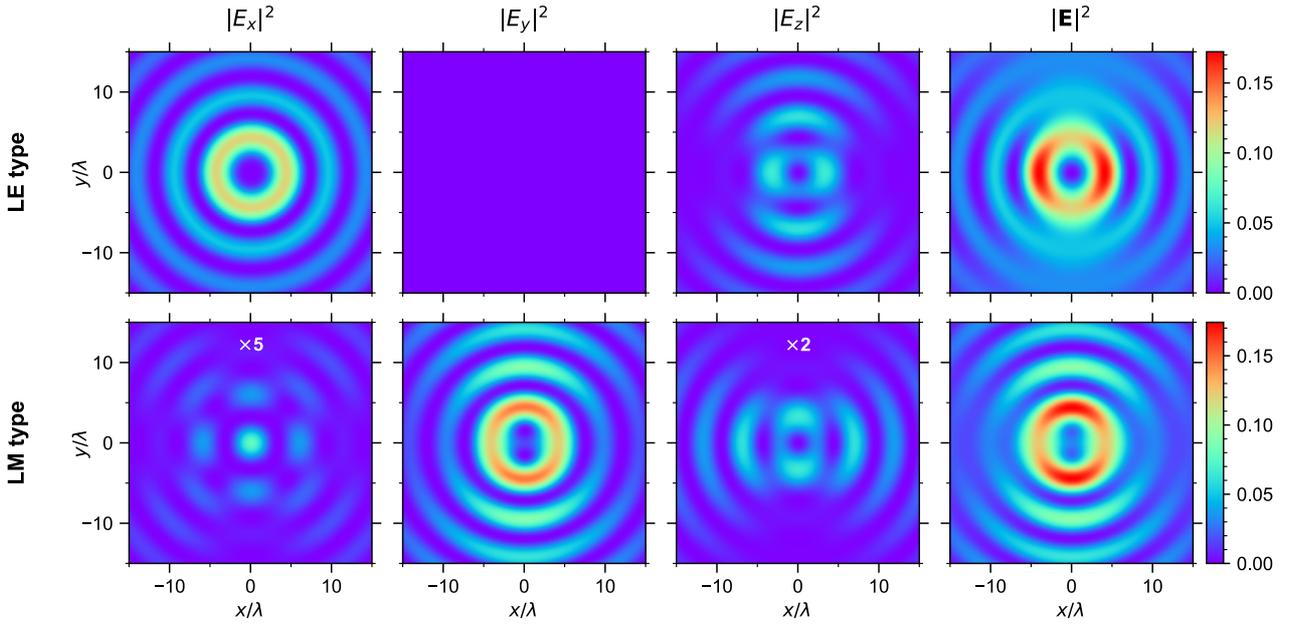

**Figure 4.** Intensity profiles of components of $\mathbf{E}_{LE}^{(x)}$ and $\mathbf{E}_{LM}^{(x)}$ with the same parameters as in Figure 2. Some components are scaled for better visibility. The $y$-component of $\mathbf{E}_{LE}^{(x)}$ is identically zero [see Eq. (21)].

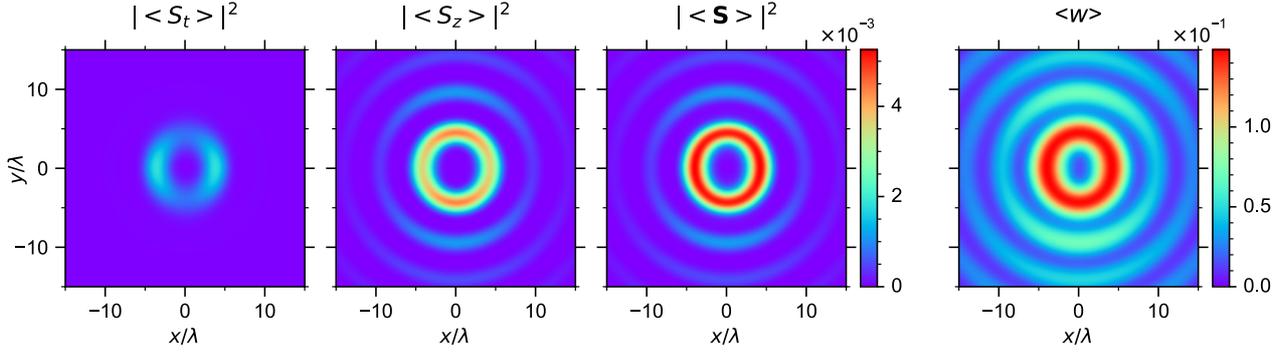

**Figure 5.** Magnitude profiles of the Poynting vector components and energy density for either $\mathbf{E}_{LE}^{(x)}$ or $\mathbf{E}_{LM}^{(x)}$ with the same parameters as in Figure 2.

Circularly symmetric (CS) Bessel beam types are defined by $\mathbf{\Pi}_m = \eta^{-1}\Pi_0 f_n \mathbf{e}_y/2$, $\mathbf{\Pi}_e = \Pi_0 f_n \mathbf{e}_x/2$ and $\mathbf{\Pi}_m = -\eta^{-1}\Pi_0 f_n \mathbf{e}_x/2$, $\mathbf{\Pi}_e = \Pi_0 f_n \mathbf{e}_y/2$ leading to two Bessel beam polarizations $\mathbf{E}_{CS}^{(1,0)}$ and $\mathbf{E}_{CS}^{(0,1)}$ (the subscript CS is usually omitted in the literature) with circularly symmetric magnitude of the Poynting vector. To make the notation of different Bessel beams more uniform, we introduce the equivalent definitions $\mathbf{E}_{CS}^{(x)} \stackrel{\text{def}}{=} \mathbf{E}_{CS}^{(1,0)}$, $\mathbf{E}_{CS}^{(y)} \stackrel{\text{def}}{=} \mathbf{E}_{CS}^{(0,1)}$, which are further justified in Section 3.2. This beam type can also be produced using the angular-spectrum representation.[17] The electric field of the CS beam is the following

$$E_{CS,x}^{(x)} = \frac{E_0}{4k^2}\left[(k+k_z)^2 f_n + \frac{k_t^2}{2}(f_{n-2} + f_{n+2})\right],$$

$$E_{CS,y}^{(x)} = i\frac{E_0 k_t^2}{8k^2}(f_{n-2} - f_{n+2}), \qquad E_{CS,z}^{(x)} = i\frac{E_0 k_t(k_z + k)}{4k^2}(f_{n-1} - f_{n+1}).$$

(23)

Magnitude profiles for the electric field, Poynting vector, and energy density of the CS beam are presented in Figure 6 and Figure 7.

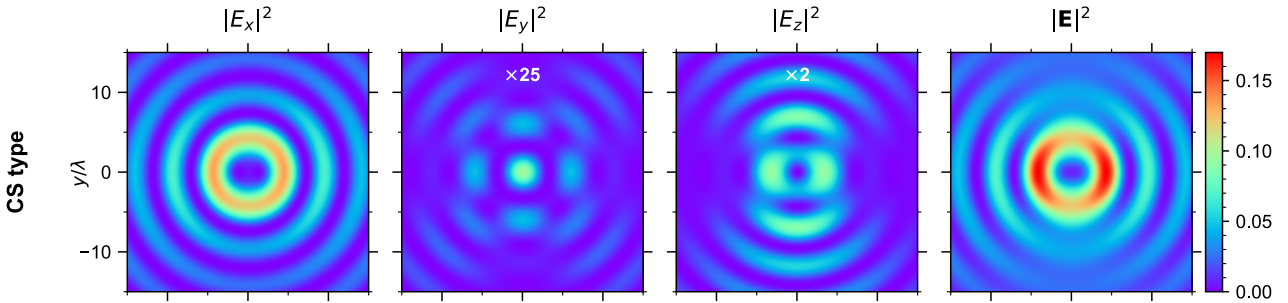

**Figure 6.** Intensity profiles of components of $\mathbf{E}_{CS}^{(x)}$ with the same parameters as in Figure 2. Some components are scaled for better visibility.

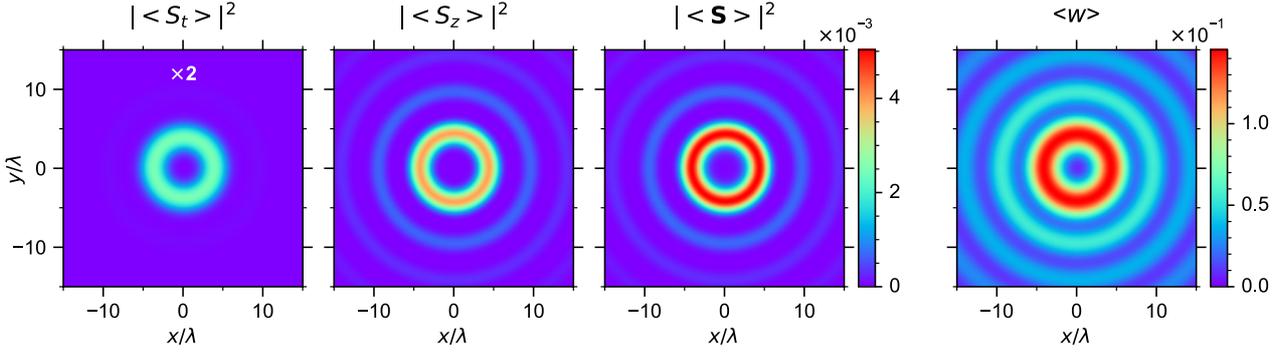

**Figure 7.** Magnitude profiles of the Poynting vector components and energy density for $\mathbf{E}_{CS}^{(x)}$ with the same parameters as in Figure 2. Some components are scaled for better visibility.

One may note that the choice of minus in one of the expressions for $\mathbf{\Pi}_m$ in the definition of the CS beam is somewhat arbitrary. Another set of CS beams can be obtained by inverting this sign (for both polarizations) – we will further denote them as CS′. We are not aware of their occurrence in the literature, but we discuss them in detail in further sections. In particular, magnitude profiles for electric field, Poynting vector, and energy density of the CS′ beam are presented in Section 3.4. Here we only provide the explicit expressions for $\mathbf{E}_{CS'}^{(x)}$, which corresponds to $\mathbf{\Pi}_m = -\eta^{-1}\Pi_0 f_n \mathbf{e}_y/2$, $\mathbf{\Pi}_e = \Pi_0 f_n \mathbf{e}_x/2$:

$$E_{CS',x}^{(x)} = \frac{E_0}{4k^2}\left[(k-k_z)^2 f_n + \frac{k_t^2}{2}(f_{n-2} + f_{n+2})\right],$$

$$E_{CS',y}^{(x)} = i\frac{E_0 k_t^2}{8k^2}(f_{n-2} - f_{n+2}), \qquad E_{CS',z}^{(x)} = i\frac{E_0 k_t(k_z - k)}{4k^2}(f_{n-1} - f_{n+1}). \tag{24}$$

## 3. Generalization of all Bessel beam types

### 3.1. Combination of transverse Hertz vector potentials

Let us first introduce a few convenient definitions. The Maxwell's equations (1) are invariant with respect to the duality transformation or, more generally, to the duality rotation expressed as[48,49]

$$\begin{pmatrix}\mathbf{E}\\\mathbf{H}\end{pmatrix} \to \mathbf{P}_\chi \begin{pmatrix}\mathbf{E}\\\mathbf{H}\end{pmatrix}, \quad \mathbf{P}_\chi \stackrel{\text{def}}{=} \begin{pmatrix}\cos\chi & -\eta\sin\chi\\ \eta^{-1}\sin\chi & \cos\chi\end{pmatrix} = \begin{pmatrix}1 & 0\\ 0 & \eta^{-1}\end{pmatrix}\mathbf{R}_\chi\begin{pmatrix}1 & 0\\ 0 & \eta\end{pmatrix}, \tag{25}$$

where $\mathbf{R}_\chi$ is the standard 2×2 rotation matrix over the angle $\chi$ (may even be complex):

$$\mathbf{R}_\chi \stackrel{\text{def}}{=} \begin{pmatrix}\cos\chi & -\sin\chi\\ \sin\chi & \cos\chi\end{pmatrix}, \tag{26}$$

which is effectively applied to the pair $(\mathbf{E}, \eta\mathbf{H})$. We also define $\mathbf{R} \stackrel{\text{def}}{=} \mathbf{R}_{\pi/2}$ for brevity. The standard duality transformations correspond to $\mathbf{P}_{\pm\pi/2}$. The duality invariance follows trivially from representation of Eq. (1) as

$$\left(ik\mathbf{P}_{\pi/2} + \nabla\times\right)\begin{pmatrix}\mathbf{E}\\ \mathbf{H}\end{pmatrix} = \begin{pmatrix}\mathbf{0}\\ \mathbf{0}\end{pmatrix}, \tag{27}$$

and noticing that transformations $\mathbf{P}_\chi$ (or rotations $\mathbf{R}_\chi$) for any complex $\chi$ commute with each other.

The same process can be applied to the pair of Hertz vector potentials. For that we write Eqs. (8), (9) in a matrix form:

$$\begin{pmatrix}\mathbf{E}\\ \mathbf{H}\end{pmatrix} = \mathcal{L}\begin{pmatrix}\mathbf{\Pi}_e\\ \mathbf{\Pi}_m\end{pmatrix}, \tag{28}$$

where $\mathcal{L}$ is a differential matrix operator:

$$\mathcal{L} \stackrel{\text{def}}{=} \nabla \times \begin{pmatrix}\nabla\times & ik\eta \\ -ik\eta^{-1} & \nabla\times\end{pmatrix} = \nabla \times (-ik\mathbf{P}_{\pi/2} + \nabla\times), \tag{29}$$

and the vector operations (like curl) evidently commute with linear superposition operations (like duality rotation). This decomposition of $\mathcal{L}$ implies that it commutes with $\mathbf{P}_\chi$, i.e. $\mathcal{L}\mathbf{P}_\chi = \mathbf{P}_\chi\mathcal{L}$, which together with Eq. (28) implies that Eq. (25) can be equivalently applied to the pair $(\mathbf{\Pi}_e, \mathbf{\Pi}_m)$ instead. Combining Eqs. (27)–(29) we obtain

$$\nabla \times (\nabla^2 + k^2)\begin{pmatrix}\mathbf{\Pi}_e\\ \mathbf{\Pi}_m\end{pmatrix} = \begin{pmatrix}\mathbf{0}\\ \mathbf{0}\end{pmatrix}, \tag{30}$$

i.e., the satisfaction of Helmholtz equation by $\mathbf{\Pi}_e$ and $\mathbf{\Pi}_m$ is a sufficient, but not necessary condition (it is necessary only for their curls), which explains why the former was postulated in Section 2.1.

Both linear (LE and LM) and circularly polarized (CS) Bessel beam types correspond to transverse Hertz potentials, i.e., those limited to the $xy$-plane. A natural generalization of these cases is a definition through an arbitrary complex matrix $\mathbf{M}$:

$$\begin{pmatrix}\mathbf{\Pi}_e\\ \mathbf{\Pi}_m\end{pmatrix} = \begin{pmatrix}\Pi_{e,x} & \Pi_{e,y}\\ \Pi_{m,x} & \Pi_{m,y}\end{pmatrix}\begin{pmatrix}\mathbf{e}_x\\ \mathbf{e}_y\end{pmatrix} = \Pi_0\begin{pmatrix}1 & 0\\ 0 & \eta^{-1}\end{pmatrix}\mathbf{M}\begin{pmatrix}\mathbf{e}_x\\ \mathbf{e}_y\end{pmatrix}f_n. \tag{31}$$

As discussed in Section 2.1, any such linear combination leads to the Hertz potentials automatically satisfying the Helmholtz equation and, hence, the resulting fields satisfy Maxwell's equations. Another convenient property of this matrix is that duality transformation $\mathbf{P}_\chi$ of Hertz potentials is equivalent to $\mathbf{M} \to \mathbf{R}_\chi\mathbf{M}$, while rotation in the $xy$-plane by the angle $\psi$ (rotation of field polarization) – to $\mathbf{M} \to \mathbf{M}\mathbf{R}_{-\psi}$ (explained in Section 3.2).

While the matrix $\mathbf{M}$ is very simple for most of existing Bessel beam types (see Section 3.3), the expressions for field components derived using Eqs. (14)–(18) are simpler using the matrix $\mathbf{M}'$, corresponding to the basis vectors $\mathbf{e}_\pm$:

$$\mathbf{M}\begin{pmatrix}\mathbf{e}_x\\ \mathbf{e}_y\end{pmatrix} = \mathbf{M}'\begin{pmatrix}\mathbf{e}_+\\ \mathbf{e}_-\end{pmatrix} \Leftrightarrow \mathbf{M} = \mathbf{M}'\mathbf{W} \Leftrightarrow \mathbf{M}' = \mathbf{M}\mathbf{W}^{-1}, \tag{32}$$

where $\mathbf{W}$ is the basis-transformation matrix:

$$\mathbf{W} = \begin{pmatrix}1 & i\\ 1 & -i\end{pmatrix}, \quad \mathbf{W}^{-1} = \frac{1}{2}\begin{pmatrix}1 & 1\\ -i & i\end{pmatrix}. \tag{33}$$

To avoid confusion, we will use different subscripts to denote components of these matrices, but omit primes for components of matrix $\mathbf{M}'$:

$$\mathbf{M} = \begin{pmatrix} M_{e,x} & M_{e,y} \\ M_{m,x} & M_{m,y} \end{pmatrix}, \qquad \mathbf{M}' = \begin{pmatrix} M_{e,+} & M_{e,-} \\ M_{m,+} & M_{m,-} \end{pmatrix}. \tag{34}$$

The general expressions for $\mathbf{E}$ in the basis $\mathbf{e}_z, \mathbf{e}_\pm$ are obtained from Eqs. (15), (18), (28)–(32):

$$E_\pm = \Pi_0 \left[ \left( \frac{k^2 + k_z^2}{2} M_{e,\pm} \pm i k k_z M_{m,\pm} \right) f_n + \frac{k_t^2}{2} M_{e,\mp} f_{n \mp 2} \right],$$

$$E_z = \Pi_0 k_t \left[ (i k_z M_{e,-} + k M_{m,-}) f_{n-1} - (i k_z M_{e,+} - k M_{m,+}) f_{n+1} \right], \tag{35}$$

which can be recast into the Descartes basis as well:

$$E_x = \Pi_0 \left\{ \left( \frac{k^2 + k_z^2}{2} M_{e,x} + k k_z M_{m,y} \right) f_n \right.$$
$$\left. + \frac{k_t^2}{4} \left[ (M_{e,x} + i M_{e,y}) f_{n-2} + (M_{e,x} - i M_{e,y}) f_{n+2} \right] \right\},$$

$$E_y = \Pi_0 \left\{ \left( \frac{k^2 + k_z^2}{2} M_{e,y} - k k_z M_{m,x} \right) f_n \right. \tag{36}$$
$$\left. + i \frac{k_t^2}{4} \left[ (M_{e,x} + i M_{e,y}) f_{n-2} - (M_{e,x} - i M_{e,y}) f_{n+2} \right] \right\},$$

$$E_z = \frac{\Pi_0 k_t}{2} \left\{ \left[ i k_z (M_{e,x} + i M_{e,y}) + k (M_{m,x} + i M_{m,y}) \right] f_{n-1} \right.$$
$$\left. - \left[ i k_z (M_{e,x} - i M_{e,y}) - k (M_{m,x} - i M_{m,y}) \right] f_{n+1} \right\}.$$

The expressions for the linearly and circularly polarized Bessel beams (LE, LM, and CS) in Section 2.2 are specific cases of Eq. (36).

At this point one may wonder why we have not included $z$-components of the potentials in Eq. (31), e.g., to describe the TE and TM Bessel beams. To explain this, let us recall that any gradient can be added to either $\mathbf{\Pi}_e$ or $\mathbf{\Pi}_m$ without changing the fields. Applying this argument to $\nabla f_n$ [Eq. (14)], we see that any potential of the form $f_n \mathbf{e}_z$ is equivalent to the superposition of transverse potentials of orders $n \pm 1$.

Let us further discuss the ambiguity of potentials. The basic functions $f_n$ for different orders $n \in \mathbb{Z}$ are linearly independent. Therefore, if all orders $n$ are considered at once, Eq. (31) has 4 (complex) degrees of freedom (matrix $\mathbf{M}$) per order (keeping $k_z$ and $k_t$ fixed). By contrast, $\mathbf{E}$ is also a linear superposition of $f_n$, but its divergence [again, superposition of $f_n$, see Eq. (16)] must be zero (one constraint for each order). Therefore, the fields have only two degrees of freedom per order, since $\mathbf{H}$ is fully determined by $\mathbf{E}$. To determine the relations between matrices $\mathbf{M}$ for various orders, note that any pair of potentials of the form

$$\left( i k \mathbf{P}_{\pi/2} + \nabla \times \right) \begin{pmatrix} \mathbf{U} \\ \mathbf{V} \end{pmatrix} - \nabla \begin{pmatrix} u \\ v \end{pmatrix}, \tag{37}$$

where $u$, $v$, $\mathbf{U}$, and $\mathbf{V}$ are arbitrary solutions to scalar and vector Helmholtz equation, respectively, is also a pair of solution to this equation and belongs to the null space of operator $\mathcal{L}$ (also mentioned in Section 2.1). In the following we express these functions as a superposition of several $f_n$.

The action of curl can be simplified using Eqs. (14), (15):

$$\nabla \times (f_n \mathbf{e}_\pm) \sim \pm \frac{1}{2k_z}[(k^2 + k_z^2)f_n \mathbf{e}_\pm - k_t^2 f_{n\pm 2}\mathbf{e}_\mp], \tag{38}$$

where the equivalence is up to a gradient. We will not track the explicit values of $u$ and $v$, noting that their values are determined by the requirement of transversality of $\mathbf{U}$ and $\mathbf{V}$. Using Eq. (38) we obtain:

$$\nabla \times \mathbf{A}\begin{pmatrix}\mathbf{e}_+\\ \mathbf{e}_-\end{pmatrix}f_n \sim \mathbf{A}\frac{1}{2k_z}\begin{pmatrix}k^2+k_z^2 & -k_t^2\mathcal{N}_2\\ k_t^2\mathcal{N}_{-2} & -(k^2+k_z^2)\end{pmatrix}\begin{pmatrix}\mathbf{e}_+\\ \mathbf{e}_-\end{pmatrix}f_n, \tag{39}$$

where the operator $\mathcal{N}$ changes the order of $f_n$ ($\mathcal{N}_l f_n \stackrel{\text{def}}{=} f_{n+l}$) and $\mathbf{A}$ is an arbitrary matrix with the same structure as $\mathbf{M}'$. Combining Eq. (39) with Eqs. (31), (37) we obtain the following trivial increment for $\mathbf{M}'$:

$$ik\mathbf{R}\mathbf{A} + \frac{k^2+k_z^2}{2k_z}\mathbf{A}\begin{pmatrix}1 & 0\\ 0 & -1\end{pmatrix} + \frac{k_t^2}{2k_z}\mathbf{A}\begin{pmatrix}0 & -\mathcal{N}_2\\ \mathcal{N}_{-2} & 0\end{pmatrix} \sim \mathbf{0}, \tag{40}$$

i.e. it can be added to $\mathbf{M}'$ without changing the fields. Denoting the columns of $\mathbf{A}$ as $\mathbf{A}_+$ and $\mathbf{A}_-$, we rewrite Eq. (40) as

$$(\mathbf{R}_\xi \mathbf{A}_+ \quad -\mathbf{R}_{-\xi}\mathbf{A}_-) \sim (-\mathcal{N}_{-2}\mathbf{A}_- \quad \mathcal{N}_2 \mathbf{A}_+), \tag{41}$$

where we introduced imaginary rotation angle $\xi$:

$$\xi \stackrel{\text{def}}{=} \text{i}\operatorname{arsinh}\frac{2kk_z}{k_t^2} = 2\text{i}\ln\frac{k+k_z}{k_t} = 2\text{i}\ln\cot\frac{\alpha_0}{2} \Leftrightarrow \mathbf{R}_\xi = \frac{1}{k_t^2}\begin{pmatrix}k^2+k_z^2 & -2\text{i}kk_z\\ 2\text{i}kk_z & k^2+k_z^2\end{pmatrix}. \tag{42}$$

Recalling that $\mathbf{A}$ is an arbitrary matrix, we equate the left-hand side of Eq. (41) to $\mathbf{M}'$, which leads to a surprisingly concise expression

$$\mathbf{M}' \sim (\mathcal{N}_{-2}\mathbf{R}_\xi \mathbf{M}'_- \quad \mathcal{N}_2 \mathbf{R}_{-\xi}\mathbf{M}'_+) = \mathcal{N}_{-2}\mathbf{R}_\xi \mathbf{M}'\begin{pmatrix}0 & 0\\ 1 & 0\end{pmatrix} + \mathcal{N}_2\mathbf{R}_{-\xi}\mathbf{M}'\begin{pmatrix}0 & 1\\ 0 & 0\end{pmatrix}. \tag{43}$$

The equivalent expressions in the Descartes basis are obtained using Eq. (32):

$$\mathbf{M} \sim \frac{1}{2}\left[\mathcal{N}_{-2}\mathbf{R}_\xi \mathbf{M}\begin{pmatrix}1 & \text{i}\\ \text{i} & -1\end{pmatrix} + \mathcal{N}_2\mathbf{R}_{-\xi}\mathbf{M}\begin{pmatrix}1 & -\text{i}\\ -\text{i} & -1\end{pmatrix}\right]. \tag{44}$$

We proved that arbitrary pair of transverse Hertz vector potentials of order $n$ is equivalent (in terms of resulting field) to a superposition of the potentials of orders $n \pm 2$. This implies that every second odd and every second even order are, strictly speaking, redundant. However, if only one order $n$ is considered, as is common in applications, then all elements of $\mathbf{M}$ (or $\mathbf{M}'$) lead to linearly independent fields.

Let us now come back to longitudinal potentials, specifically to a general expression of order $n$, specified by 2D vector of coefficients $\mathbf{q}$:

$$\begin{pmatrix} \mathbf{\Pi}_e \\ \mathbf{\Pi}_m \end{pmatrix} = \Pi_0 \begin{pmatrix} 1 & 0 \\ 0 & \eta^{-1} \end{pmatrix} \begin{pmatrix} q_e \\ q_m \end{pmatrix} f_n \mathbf{e}_z. \tag{45}$$

Using Eqs. (15), (43) the equivalent transverse potentials can be expressed as

$$\mathbf{M}' = \frac{ik_t}{2k_z}(\mathcal{N}_{-1}\mathbf{q} \quad -\mathcal{N}_1\mathbf{q}) \sim \frac{ik_t}{2k_z}(\mathbf{1} - \mathbf{R}_\xi)(\mathbf{q} \quad \mathbf{0})\mathcal{N}_{-1} \sim -\frac{ik_t}{2k_z}(\mathbf{1} - \mathbf{R}_{-\xi})(\mathbf{0} \quad \mathbf{q})\mathcal{N}_1. \tag{46}$$

Importantly, it is sufficient to use transverse potentials of a single order (either $n - 1$ or $n + 1$) and only a single direction ($\mathbf{e}_+$ or $\mathbf{e}_-$, respectively), but at a cost of using both $\mathbf{\Pi}_e$ and $\mathbf{\Pi}_m$ even for the simplest TE and TM beams (see specific expressions in Section 3.3). Conversely, Eq. (46) implies that any transverse potential of order $n$ is equivalent to a superposition of longitudinal potentials of orders $n \pm 1$. Therefore, the set of longitudinal Hertz vector potentials of all orders is another complete basis for Bessel beams, equivalent to the basis of CVWFs.

To conclude this section, the above results show that specifying the matrix $\mathbf{M}$ or $\mathbf{M}'$ for a specific order $n$ allows one not only to describe the Bessel beams for arbitrary transverse Hertz vector potential of order $n$, but also for arbitrary longitudinal potentials of orders $n \pm 1$, for transverse potentials of orders $n \pm 2$ proportional to $\mathbf{e}_-$ and $\mathbf{e}_+$, respectively, and for any linear combination of the above.

### 3.2. Polarization rotation

Most light-scattering codes are tailored for the calculation of the Mueller (or amplitude) scattering matrices, which requires simulations for two polarizations (commonly linear) of the incident field.[23] These polarizations need to be connected by $\pi/2$ rotation to enable rotation relations for the scattering matrices (discussed below in Section 4.1). In the case of Bessel beams (of any specific type), our goal is to also define two such basis polarizations:

$$\mathbf{E}^\parallel(\mathbf{r}) \propto \mathcal{R}_{\pi/2} \mathbf{E}^\perp(\mathbf{r}) = \widetilde{\mathbf{R}}_{\pi/2} \mathbf{E}^\perp(\widetilde{\mathbf{R}}_{-\pi/2}\mathbf{r}), \tag{47}$$

where $\mathcal{R}_\psi$ is the rotation operator (acting on a field) and $\widetilde{\mathbf{R}}_\psi$ is the 3×3 rotation matrix (acting on a vector) over the angle $\chi$ around the beam propagation axis (positive value – in a counterclockwise direction). For the default propagation along the $z$-axis (as considered in this manuscript), $\widetilde{\mathbf{R}}_\psi$ is exactly 2D-rotation matrix $\mathbf{R}_\psi$ accompanied by 1 at the position (3,3). Parallel and perpendicular polarizations are considered with respect to the scattering plane, as typically used for scattering matrices.[23] For a plane wave the rotation of $\mathbf{r}$ in Eq. (47) is redundant, and the proportionality can be replaced by equality. By contrast, a Bessel beam is generally a vortex beam (i.e., its phase depends on the azimuthal angle $\varphi$) leading to the additional phase factor discussed below.

Using the definitions of Section 3.1 [Eqs. (28), (31)] we can apply rotation transformation to arbitrary Bessel beam, for both electric and magnetic fields:

$$\mathcal{R}_\psi \begin{pmatrix} \mathbf{E}(\mathbf{r}) \\ \mathbf{H}(\mathbf{r}) \end{pmatrix} = \begin{pmatrix} \widetilde{\mathbf{R}}_\psi \mathbf{E}(\widetilde{\mathbf{R}}_{-\psi}\mathbf{r}) \\ \widetilde{\mathbf{R}}_\psi \mathbf{H}(\widetilde{\mathbf{R}}_{-\psi}\mathbf{r}) \end{pmatrix} = \Pi_0 \mathcal{L} \begin{pmatrix} 1 & 0 \\ 0 & \eta^{-1} \end{pmatrix} \mathbf{M}\mathbf{R}_{-\psi} \begin{pmatrix} \mathbf{e}_x \\ \mathbf{e}_y \end{pmatrix} f_n(\widetilde{\mathbf{R}}_{-\psi}\mathbf{r}), \tag{48}$$

where we used that the rotation and curl operators commute, rotation of each element in a row is equivalent to multiplication by $\widetilde{\mathbf{R}}_\psi^\mathrm{T} = \widetilde{\mathbf{R}}_{-\psi}$ from the right, and

$$\begin{pmatrix} \mathbf{e}_x \\ \mathbf{e}_y \end{pmatrix} \widetilde{\mathbf{R}}_{-\psi} = \mathbf{R}_{-\psi} \begin{pmatrix} \mathbf{e}_x \\ \mathbf{e}_y \end{pmatrix}. \tag{49}$$

Moreover, the rotation of the argument of $f_n$ is trivial:

$$f_n(\widetilde{\mathbf{R}}_{-\psi}\mathbf{r}) = f_n(\rho, \varphi - \psi, z) = e^{-in\psi} f_n(\mathbf{r}), \tag{50}$$

leading to the following general relation:

$$\mathcal{R}_\psi \mathbf{E}_n(\mathbf{r}, \mathbf{M}) = \mathbf{E}_n(\mathbf{r}, e^{-in\psi} \mathbf{M} \mathbf{R}_{-\psi}). \tag{51}$$

In other words, the rotation of any Bessel beam is equivalent to the transformation of its defining matrix $\mathbf{M}$. In particular, the rotations through the angles $\pi$ and $2\pi$ are especially simple:

$$\mathcal{R}_\pi \mathbf{E}_n(\mathbf{r}, \mathbf{M}) = (-1)^{n+1} \mathbf{E}_n(\mathbf{r}, \mathbf{M}), \qquad \mathcal{R}_{2\pi} \mathbf{E}_n(\mathbf{r}, \mathbf{M}) = \mathbf{E}_n(\mathbf{r}, \mathbf{M}). \tag{52}$$

Here and further in this section we discuss only electric fields, since the corresponding rotation expressions for magnetic fields are the same.

Based on the above, we postulate the two orthogonal polarizations of the Bessel beams to be related as [cf. Eq. (47)]

$$\mathcal{R}_{\pi/2} \mathbf{E}^\perp = i^{-n} \mathbf{E}^\parallel \Leftrightarrow \mathbf{M}_\perp = \mathbf{M}_\parallel \mathbf{R}, \tag{53}$$

i.e., their matrices are related by simple rotation without any additional phase factors. We discuss the exact meaning of orthogonality in Section 3.5. Here we note, that this definition complies with existing linear and circular polarizations of the Bessel beams, discussed in Section 2.2 (corresponding to the rotation of Hertz vector potentials). In particular, the pair $\{\mathbf{E}_{\ldots}^{(x)}, \mathbf{E}_{\ldots}^{(y)}\}$ for any of LE, LM, CS, and CS′ beams can be used as a pair $\{\mathbf{E}^\perp, \mathbf{E}^\parallel\}$, although their relation to the scattering plane ($yz$-plane by default) cannot be described by simple "perpendicular" or "parallel" notions.

The only missing component is the polarizations for TE and TM types, since they are almost axisymmetric and have trivial rotation transformations (due to their expressions through the $z$-polarized Hertz vector potentials):

$$\mathcal{R}_\psi \mathbf{E}_{\mathrm{TE,TM},n} = e^{-in\psi} \mathbf{E}_{\mathrm{TE,TM},n}, \tag{54}$$

where we explicitly specify the order of Bessel beam as a subscript to avoid confusion. Let us introduce $x$- and $y$-components for these beams ($\mathbf{E}_{\mathrm{TEL}}^{(x)}$, $\mathbf{E}_{\mathrm{TEL}}^{(y)}$, $\mathbf{E}_{\mathrm{TML}}^{(x)}$, $\mathbf{E}_{\mathrm{TML}}^{(y)}$) through the following relations:

$$\mathbf{E}_{\mathrm{TE,TM},n+1} = \mathbf{E}_{\mathrm{TEL,TML},n}^{(x)} + i\, \mathbf{E}_{\mathrm{TEL,TML},n}^{(y)}, \tag{55}$$

$$\mathcal{R}_{\pi/2} \mathbf{E}_{\mathrm{TEL,TML},n}^{(x)} = i^{-n} \mathbf{E}_{\mathrm{TEL,TML},n}^{(y)}, \tag{56}$$

These are not sufficient for unique determination of the new beam types. However, if we additionally require the corresponding matrices **M** to be real (for real $\alpha_0$), which is important for orthogonality relations – see Section 3.5, then the only possibility is

$$\mathbf{E}_{\text{TEL}}^{(x),(y)} \stackrel{\text{def}}{=} k_t^{-1} \left( k_z \mathbf{E}_{\text{LE}}^{(x),(y)} \pm k \mathbf{E}_{\text{LM}}^{(y),(x)} \right), \tag{57}$$

$$\mathbf{E}_{\text{TML}}^{(x),(y)} \stackrel{\text{def}}{=} k_t^{-1} \left( k_z \mathbf{E}_{\text{LM}}^{(x),(y)} \mp k \mathbf{E}_{\text{LE}}^{(y),(x)} \right), \tag{58}$$

where $\pm$ corresponds to $x$- and $y$-polarizations on the left-hand sides, respectively. The examples of the corresponding magnitude profiles for electric field, Poynting vector, and energy density are presented in Figure 8 and Figure 9.

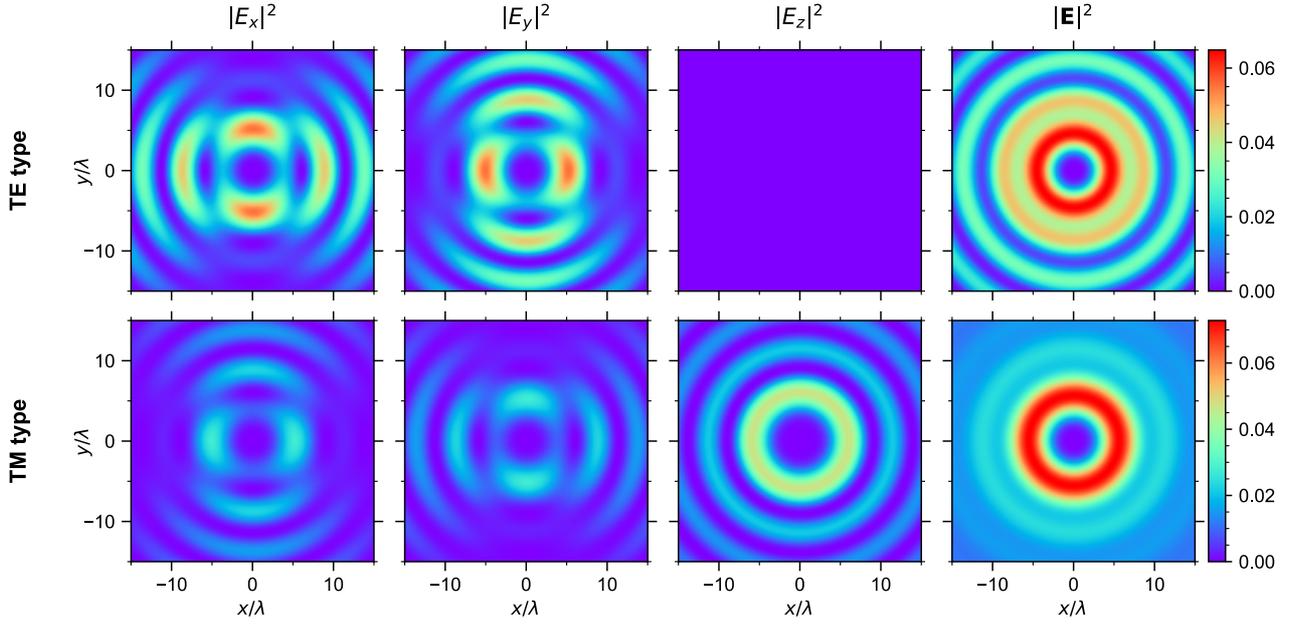

**Figure 8.** Intensity profiles of components of $\mathbf{E}_{\text{TEL}}^{(x)}$ and $\mathbf{E}_{\text{TML}}^{(x)}$ with the same parameters as in Figure 2. The $x$-component of $\mathbf{E}_{\text{TML}}^{(x)}$ is scaled for better visibility, while the $z$-component of $\mathbf{E}_{\text{TEL}}^{(x)}$ is identically zero [see Eq. (57)].

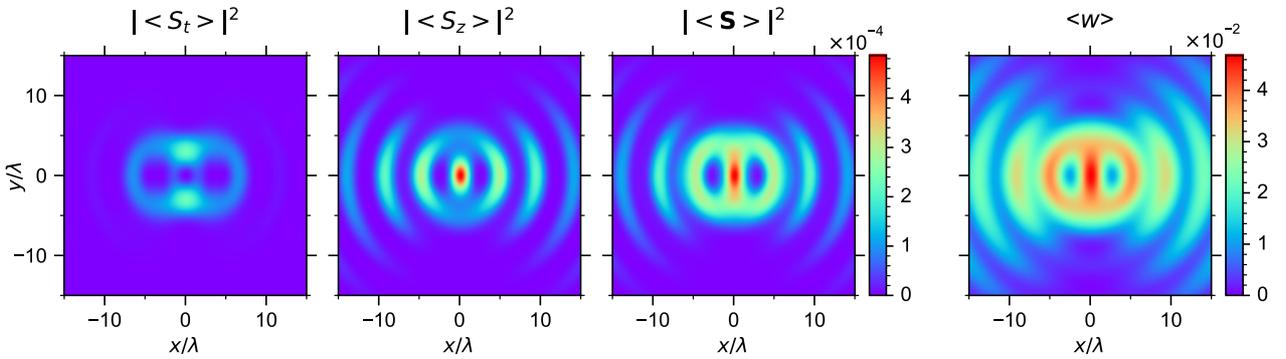

**Figure 9.** Magnitude profiles of the Poynting vector components and energy density for either $\mathbf{E}_{\text{TEL}}^{(x)}$ or $\mathbf{E}_{\text{TML}}^{(x)}$ with the same parameters as in Figure 2.

While the resulting relations between the LE, LM, TE, and TM Bessel beams are known in the literature,[18] Eqs. (54)–(58) can be trivially derived from the **M**-matrix representations of these beams (see Table 3), while the fundamental reason for the change of order between the original TE and TM beams and their components is explained in Section 3.1 [Eq. (46)]. Moreover, Eq. (56) implies that the $x$- and $y$-components form a proper pair $\{\mathbf{E}^\perp, \mathbf{E}^\parallel\}$, as was our intention.

To get better understanding of the newly defined beams, let us consider the plane wave limit, i.e. $\alpha_0 = 0°$ and $n = 0$ (for fixed **r**). The limiting expressions for these and other Bessel beam types are given in Table 1, they follow from the expressions in Section 2.2. In particular, the LE beam corresponds to the same-polarized electric field, LM corresponds to the same-polarized magnetic field and, hence, to the orthogonal (rotated) electric-field polarization. The CS beam approaches a linearly polarized plane wave, which illustrates that CS is related not to circular polarization but rather to a circularly symmetric intensity (which is always the case for any polarization of a plane wave). By contrast, the CS′ beam vanish in this limit for any order $n$ as $\mathcal{O}(\alpha_0^4)$ due to the cancelation of the corresponding contributions from electric and magnetic vector potentials. However, it is $\mathcal{O}(\alpha_0^3)$ for $n = 1$, while for $n = 2$ we have $\lim_{\alpha_0 \to 0} \mathbf{E}_{\text{CS}'}^{(x)} = \alpha_0^2 E_0 \mathbf{e}_+ / 8$, i.e. the CS′ beam is distantly related to the circular polarization. Similar limiting behavior is obtained for the TE and TM beams. They are $\mathcal{O}(\alpha_0^2)$ for $n = 0$, while

$$n = 1 \Rightarrow \lim_{\alpha_0 \to 0} \mathbf{E}_{\text{TE}} = -\alpha_0 E_0 \mathbf{e}_+/2, \quad \lim_{\alpha_0 \to 0} \mathbf{E}_{\text{TM}} = i\alpha_0 E_0 \mathbf{e}_+/2, \tag{59}$$

which follows from the limiting expressions for the TEL and TML beams of order 0. The latter are similar to the LE and LM beams, but are proportional to $\alpha_0$. Thus, the TEL and TML beams are analogous to linear polarizations of a plane wave, while the TE and TM beams – to a circular one.

**Table 1.** The electric field in two limiting cases for different Bessel beam types. Definitions of the constituent functions are given in the text. Expressions for other polarizations can be obtained by rotation [Eq. (53) and (56)].

| Type | Description | Field | Plane-wave limit, $\times E_0 e^{ikz}$ | Bullet limit, $\times E_0$ |
|------|-------------|-------|----------------------------------------|----------------------------|
| LE | Linearly polarized electric field | $\mathbf{E}_{\text{LE}}^{(x)}$ | $\mathbf{e}_x$ | $i(\mathbf{F}_{n-1}^z - \mathbf{F}_{n+1}^z)$ |
| LM | Linearly polarized magnetic field | $\mathbf{E}_{\text{LM}}^{(x)}$ | $\mathbf{e}_y$ | $i(\mathbf{F}_{n-1}^t - \mathbf{F}_{n+1}^t)$ |
| CS | Circularly symmetric energy density | $\mathbf{E}_{\text{CS}}^{(x)}$ | $\mathbf{e}_x$ | $\frac{1}{2}(\mathbf{F}_{n-1}^t + \mathbf{F}_{n+1}^t + i\mathbf{F}_{n-1}^z - i\mathbf{F}_{n+1}^z)$ |
| CS′ | Alternative CS | $\mathbf{E}_{\mathrm{CS'}}^{(x)}$ | $\mathcal{O}(\alpha_0^4)$ | $\frac{1}{2}(\mathbf{F}_{n-1}^{\mathrm{t}} + \mathbf{F}_{n+1}^{\mathrm{t}} + i\mathbf{F}_{n+1}^z - i\mathbf{F}_{n-1}^z)$ |
| TE | Transverse electric field | $\mathbf{E}_{\mathrm{TE}}$ | $\mathcal{O}(\alpha_0^2)$ | $-2\mathbf{F}_n^{\mathrm{t}}$ |
| TM | Transverse magnetic field | $\mathbf{E}_{\mathrm{TM}}$ | $\mathcal{O}(\alpha_0^2)$ | $2\mathbf{F}_n^z$ |
| TEL | Linear component of TE | $\mathbf{E}_{\mathrm{TEL}}^{(x)}$ | $-\alpha_0 \mathbf{e}_x/2$ | $-\mathbf{F}_{n-1}^{\mathrm{t}} - \mathbf{F}_{n+1}^{\mathrm{t}}$ |
| TML | Linear component of TM | $\mathbf{E}_{\mathrm{TML}}^{(x)}$ | $-\alpha_0 \mathbf{e}_y/2$ | $\mathbf{F}_{n-1}^z + \mathbf{F}_{n+1}^z$ |

Another illustrative limit, considered in Table 1, is that of a tightly focused Bessel beam, i.e. $\alpha_0 \to \pi/2$ ($k_z \to 0$, $k_t \to k$). When combined with short pulses, this limit is also known as the electromagnetic bullet.[50] The corresponding expressions make use of the following shorthand definitions:

$$\mathbf{F}_n^z \stackrel{\mathrm{def}}{=} \frac{1}{2} f_n \mathbf{e}_z, \qquad \mathbf{F}_n^{\mathrm{t}} \stackrel{\mathrm{def}}{=} \frac{1}{4}(f_{n-1}\mathbf{e}_+ + f_{n+1}\mathbf{e}_-), \qquad (60)$$

and $f_n(\mathbf{r}) \xrightarrow[\alpha_0 \to \pi/2]{} J_n(k\rho)e^{in\varphi}$. Note that both $\mathbf{F}_n^z$ and $\mathbf{F}_n^{\mathrm{t}}$ are eigenfunctions of the rotation operator:

$$\mathcal{R}_\psi \mathbf{F}_n^z = e^{-in\psi}\mathbf{F}_n^z, \qquad \mathcal{R}_\psi \mathbf{F}_n^{\mathrm{t}} = e^{-in\psi}\mathbf{F}_n^{\mathrm{t}}, \qquad (61)$$

since $\widetilde{\mathbf{R}}_\psi \mathbf{e}_\pm = e^{\mp i\psi} \mathbf{e}_\pm$ [cf. Eqs. (50), (54)]. One can see that the resulting expressions in the two considered limit are qualitatively different, in some cases transverse field is changed into a longitudinal one. That is not surprising given the complicated general expressions of the fields for intermediate $\alpha_0$.

Besides the polarizations discussed in Section 2.2, there are known composites, in particular, circular polarizations for the LE, LM, and CS Bessel beams denoted by $(1, \pm i)$ superscript.[17] Let us make an obvious generalization of this definition for $\alpha, \beta \in \mathbb{C}$:

$$\mathbf{E}_{\cdots}^{(\alpha,\beta)} \stackrel{\mathrm{def}}{=} \alpha \mathbf{E}_{\cdots}^{(x)} + \beta \mathbf{E}_{\cdots}^{(y)}, \qquad (62)$$

which applies to any pair of $\{\mathbf{E}^\perp, \mathbf{E}^\|\}$, including the LE, LM, CS, CS′, TEL, and TML beams. This definition constitutes a rigorous relation between $(x)$, $(y)$ and $(1,0)$, $(0,1)$ superscripts, respectively, used for various Bessel beams in the literature. In particular, $\mathbf{E}_{\mathrm{LE,LM}}^{(1,\pm i)}$ can be considered as generalizations of circularly-polarized plane waves, while $\alpha, \beta \in \mathbb{R}$ correspond to the rotated linear polarizations of a plane wave (up to a constant factor). In particular, the latter analogy applies to any $x$- or $y$-polarizations of these beams. Moreover, Eq. (55) can be rewritten as

$$\mathbf{E}_{\mathrm{TE},n\pm 1} = \mathbf{E}_{\mathrm{TEL},n}^{(1,\pm i)}, \qquad \mathbf{E}_{\mathrm{TM},n\pm 1} = \mathbf{E}_{\mathrm{TML},n}^{(1,\pm i)}, \qquad (63)$$

where the equalities for $n-1$ follows from Eqs. (44) or (46).

In terms of matrix $\mathbf{M}$, Eq. (62) can be represented as

$$\mathbf{M}^{(\alpha,\beta)} = \alpha \mathbf{M}^{(1,0)} + \beta \mathbf{M}^{(0,1)} = \mathbf{M}^{(1,0)} \begin{pmatrix} \alpha & \beta \\ -\beta & \alpha \end{pmatrix}, \tag{64}$$

using Eq. (53), which leads to the expressions for various beam types, summarized in Table 2. Note that both CS and CS′ beams are described by scaled (complex) orthogonal matrices, i.e.

$$\mathbf{M}_{CS}^{(\alpha,\beta)} \mathbf{M}_{CS}^{(\alpha,\beta)T} = \mathbf{M}_{CS'}^{(\alpha,\beta)} \mathbf{M}_{CS'}^{(\alpha,\beta)T} = \frac{\alpha^2 + \beta^2}{4} \mathbf{I}, \tag{65}$$

where $\mathbf{I}$ is the identity matrix. However, $\mathbf{M}_{CS}^{(\alpha,\beta)}$ is skew-symmetric (scaled rotation matrix), while $\mathbf{M}_{CS'}^{(\alpha,\beta)}$ is symmetric (scaled reflection matrix).

**Table 2.** Matrices $\mathbf{M}$ for generalized polarizations of different Bessel beam types.

| Type | $\mathbf{M}$ |
|---|---|
| $\mathbf{E}_{LE}^{(\alpha,\beta)}$ | $\begin{pmatrix} 0 & 0 \\ -\beta & \alpha \end{pmatrix}$ |
| $\mathbf{E}_{LM}^{(\alpha,\beta)}$ | $\begin{pmatrix} -\beta & \alpha \\ 0 & 0 \end{pmatrix}$ |
| $\mathbf{E}_{CS}^{(\alpha,\beta)}$ | $\frac{1}{2}\begin{pmatrix} \alpha & \beta \\ -\beta & \alpha \end{pmatrix}$ |
| $\mathbf{E}_{CS'}^{(\alpha,\beta)}$ | $\frac{1}{2}\begin{pmatrix} \alpha & \beta \\ \beta & -\alpha \end{pmatrix}$ |
| $\mathbf{E}_{TEL}^{(\alpha,\beta)}$ | $\frac{1}{k_t}\begin{pmatrix} -k\alpha & -k\beta \\ -k_z\beta & k_z\alpha \end{pmatrix}$ |
| $\mathbf{E}_{TML}^{(\alpha,\beta)}$ | $\frac{1}{k_t}\begin{pmatrix} -k_z\beta & k_z\alpha \\ k\alpha & k\beta \end{pmatrix}$ |

While there is certain redundancy in this definition (multiplication of both $\alpha$ and $\beta$ by the same factor leads to the trivial scaling of the fields) we do not postulate any specific normalization. An obvious choice for the latter when $\alpha, \beta \in \mathbb{R}$ is

$$\alpha^2 + \beta^2 = 1, \tag{66}$$

then the transformation matrix in Eq. (64) is exactly the rotation one. In the case of general complex coefficients, Eq. (66) is also an option leading to rotation matrix for a complex angle. But this normalization is inconvenient, e.g., for the abovementioned case $(1, \pm i)$, since the corresponding rotation angle is then infinite:

$$\lim_{x \to \infty} 2e^{-x} \mathbf{R}_{\mp ix} = \begin{pmatrix} 1 & \pm i \\ \mp i & 1 \end{pmatrix}. \tag{67}$$

Alternative normalization is

$$|\alpha|^2 + |\beta|^2 = 1, \tag{68}$$

which preserves the norm of the electric field (see Section 3.5).

Importantly, the generalized polarizations have simple rotation relation for arbitrary complex angle $\psi$:

$$\mathcal{R}_\psi \mathbf{E}_n^{(\alpha,\beta)} = e^{-in\psi}\left(\cos\psi\, \mathbf{E}_n^{(\alpha,\beta)} + \sin\psi\, \mathbf{E}_n^{(-\beta,\alpha)}\right), \qquad (69)$$

for arbitrary $(\alpha,\beta)$ and each Bessel beam type (subscripts LE, LM, CS, CS′, TEL, and TML). This relation follows from the matrix representation and Eqs. (53), (64). The result is a straightforward generalization of rotation relation for plane waves.[51] Moreover, note that by definition $\mathbf{E}_n^{(-1,0)} = -\mathbf{E}_n^{(1,0)}$ and Eq. (69) implies Eq. (52) for $\mathbf{E}_n^{(\alpha,\beta)}$. Thus, any pair $\mathbf{E}_n^{(\alpha,\beta)}$ and $\mathbf{E}_n^{(-\beta,\alpha)}$ can be used as a pair $\{\mathbf{E}^\perp, \mathbf{E}^\parallel\}$ [cf. Eq. (53)], but becomes degenerate whenever $(-\beta,\alpha)$ is proportional to $(\alpha,\beta)$, which is equivalent to $\alpha^2 + \beta^2 = 0$. The latter case corresponds to generalized circular polarizations $(1,\pm i)$ (up to a constant factor), which break the normalization given by Eq. (66) and are eigenfunctions of the rotation operator:

$$\mathcal{R}_\psi \mathbf{E}_n^{(1,\pm i)} = e^{-i(n\pm 1)\psi} \mathbf{E}_n^{(1,\pm i)}. \qquad (70)$$

This equation is a generalization of expressions for TE and TM beams [cf. Eqs. (54), (63)] and is simpler if written in terms of matrix $\mathbf{M}$, using Eq. (51):

$$\mathbf{M}^{(1,\pm i)} \mathbf{R}_\psi = e^{\pm i\psi} \mathbf{M}^{(1,\pm i)}. \qquad (71)$$

### 3.3. Duality rotation and relations between Bessel beam types

Duality transformations are naturally incorporated into the general framework described in Section 3.1. In particular, Table 3 demonstrates that the LE and LM, TE and TM, TEL and TML Bessel beam types are paired by the duality transformation. We do not show the corresponding expressions for matrix $\mathbf{M}'$, since they are trivially obtained from Eq. (32) and are in most cases more complicated. The only exception is the TE and TM beams, for which

$$\mathbf{M}'_{\text{TE}} = \frac{1}{k_t}\begin{pmatrix} -k & 0 \\ -ik_z & 0 \end{pmatrix}, \qquad \mathbf{M}'_{\text{TM}} = \frac{1}{k_t}\begin{pmatrix} -ik_z & 0 \\ k & 0 \end{pmatrix}. \qquad (72)$$

where the matrix $\mathbf{M}'$ of order $n$ corresponds to the TE and TM beams of order $n+1$. Eq. (72) follows from the definition of these beams in Section 2.2 and Eq. (46).

**Table 3.** The relations between different Bessel beams via polarization and duality rotations. Note that the order of the TE and TM beams is larger by 1 than the order of the corresponding matrix $\mathbf{M}$.

| Type | Field | M | Polarization rotation | Duality rotation |
|---|---|---|---|---|
| LE | $\mathbf{E}_{\text{LE}}^{(x)}$ | $\begin{pmatrix} 0 & 0 \\ 0 & 1 \end{pmatrix}$ | $\mathbf{M}_{\text{LE}}^{(y)} \mathbf{R} = \mathbf{M}_{\text{LE}}^{(x)}$ | $\mathbf{R}\mathbf{M}_{\text{LM}}^{(x,y)} = \mathbf{M}_{\text{LE}}^{(x,y)}$ |
|  | $\mathbf{E}_{\text{LE}}^{(y)}$ | $\begin{pmatrix} 0 & 0 \\ -1 & 0 \end{pmatrix}$ |  |  |
| LM | $\mathbf{E}_{\mathrm{LM}}^{(x)}$ | $\begin{pmatrix} 0 & 1 \\ 0 & 0 \end{pmatrix}$ | $\mathbf{M}_{\mathrm{LM}}^{(y)}\mathbf{R} = \mathbf{M}_{\mathrm{LM}}^{(x)}$ | |
| | $\mathbf{E}_{\mathrm{LM}}^{(y)}$ | $\begin{pmatrix} -1 & 0 \\ 0 & 0 \end{pmatrix}$ | | |
| CS | $\mathbf{E}_{\mathrm{CS}}^{(x)}$ | $\frac{1}{2}\begin{pmatrix} 1 & 0 \\ 0 & 1 \end{pmatrix}$ | $\mathbf{M}_{\mathrm{CS}}^{(y)}\mathbf{R} = \mathbf{M}_{\mathrm{CS}}^{(x)}$ | $\mathbf{R}\mathbf{M}_{\mathrm{CS}}^{(y)} = \mathbf{M}_{\mathrm{CS}}^{(x)}$ |
| | $\mathbf{E}_{\mathrm{CS}}^{(y)}$ | $\frac{1}{2}\begin{pmatrix} 0 & 1 \\ -1 & 0 \end{pmatrix}$ | | |
| CS′ | $\mathbf{E}_{\mathrm{CS'}}^{(x)}$ | $\frac{1}{2}\begin{pmatrix} 1 & 0 \\ 0 & -1 \end{pmatrix}$ | $\mathbf{M}_{\mathrm{CS'}}^{(y)}\mathbf{R} = \mathbf{M}_{\mathrm{CS'}}^{(x)}$ | $\mathbf{R}\mathbf{M}_{\mathrm{CS'}}^{(x)} = \mathbf{M}_{\mathrm{CS'}}^{(y)}$ |
| | $\mathbf{E}_{\mathrm{CS'}}^{(y)}$ | $\frac{1}{2}\begin{pmatrix} 0 & 1 \\ 1 & 0 \end{pmatrix}$ | | |
| TE | $\mathbf{E}_{\mathrm{TE},n+1}$ | $\frac{1}{k_t}\begin{pmatrix} -k & -ik \\ -ik_z & k_z \end{pmatrix}$ | $\mathbf{M}_{\mathrm{TE}}\mathbf{R}_\psi = e^{i\psi}\mathbf{M}_{\mathrm{TE}}$ | $\mathbf{R}\mathbf{M}_{\mathrm{TM}} = \mathbf{M}_{\mathrm{TE}}$ |
| TM | $\mathbf{E}_{\mathrm{TM},n+1}$ | $\frac{1}{k_t}\begin{pmatrix} -ik_z & k_z \\ k & ik \end{pmatrix}$ | $\mathbf{M}_{\mathrm{TM}}\mathbf{R}_\psi = e^{i\psi}\mathbf{M}_{\mathrm{TM}}$ | |
| TEL | $\mathbf{E}_{\mathrm{TEL}}^{(x)}$ | $\frac{1}{k_t}\begin{pmatrix} -k & 0 \\ 0 & k_z \end{pmatrix}$ | $\mathbf{M}_{\mathrm{TEL}}^{(y)}\mathbf{R} = \mathbf{M}_{\mathrm{TEL}}^{(x)}$ | $\mathbf{R}\mathbf{M}_{\mathrm{TML}}^{(x,y)} = \mathbf{M}_{\mathrm{TEL}}^{(x,y)}$ |
| | $\mathbf{E}_{\mathrm{TEL}}^{(y)}$ | $\frac{1}{k_t}\begin{pmatrix} 0 & -k \\ -k_z & 0 \end{pmatrix}$ | | |
| TML | $\mathbf{E}_{\mathrm{TML}}^{(x)}$ | $\frac{1}{k_t}\begin{pmatrix} 0 & k_z \\ k & 0 \end{pmatrix}$ | $\mathbf{M}_{\mathrm{TML}}^{(y)}\mathbf{R} = \mathbf{M}_{\mathrm{TML}}^{(x)}$ | |
| | $\mathbf{E}_{\mathrm{TML}}^{(y)}$ | $\frac{1}{k_t}\begin{pmatrix} -k_z & 0 \\ 0 & k \end{pmatrix}$ | | |

For the CS beams the matrix $\mathbf{M}$ is a (scaled) rotation matrix itself, even for $(\alpha, \beta)$ – see Table 2, hence it commutes with other rotations:

$$\mathbf{R}_\psi \mathbf{M}_{\mathrm{CS}}^{(\alpha,\beta)} = \mathbf{M}_{\mathrm{CS}}^{(\alpha,\beta)} \mathbf{R}_\psi, \tag{73}$$

That means that a duality rotation is equivalent to the polarization rotation by the inverse angle. By contrast, for the CS′ beams the matrix $\mathbf{M}$ is a (scaled) reflection matrix, thus it satisfies

$$\mathbf{R}_\psi \mathbf{M}_{\mathrm{CS'}}^{(\alpha,\beta)} = \mathbf{M}_{\mathrm{CS'}}^{(\alpha,\beta)} \mathbf{R}_{-\psi}. \tag{74}$$

Moreover, the symmetry of the CS beams under polarization and duality rotation leads to the symmetry of the time-averaged energy density and magnitude of the Poynting vector that will be discussed in Section 3.4.

Another important property, evident from Table 3, is that a beam defined by an arbitrary matrix **M** is a linear combination of $\mathbf{E}_{LE}^{(x)}, \mathbf{E}_{LE}^{(y)}, \mathbf{E}_{LM}^{(x)}, \mathbf{E}_{LM}^{(y)}$. Therefore, one can immediately express other Bessel beam types through these 4 beams [see also Eqs. (57),(58)]:

$$\mathbf{E}_{CS}^{(x),(y)} = \left(\mathbf{E}_{LE}^{(x),(y)} \mp \mathbf{E}_{LM}^{(y),(x)}\right)/2, \tag{75}$$

$$\mathbf{E}_{CS'}^{(x),(y)} = \left(-\mathbf{E}_{LE}^{(x),(y)} \mp \mathbf{E}_{LM}^{(y),(x)}\right)/2, \tag{76}$$

$$\mathbf{E}_{TE,n+1} = k_t^{-1}\left[-ik\left(\mathbf{E}_{LM,n}^{(x)} + i\mathbf{E}_{LM,n}^{(y)}\right) + k_z\left(\mathbf{E}_{LE,n}^{(x)} + i\mathbf{E}_{LE,n}^{(y)}\right)\right], \tag{77}$$

$$\mathbf{E}_{TM,n+1} = k_t^{-1}\left[ik\left(\mathbf{E}_{LE,n}^{(x)} + i\mathbf{E}_{LE,n}^{(y)}\right) + k_z\left(\mathbf{E}_{LM,n}^{(x)} + i\mathbf{E}_{LM,n}^{(y)}\right)\right], \tag{78}$$

where the beam orders are specified only when they are different inside one equation.

Naturally, there are other ways to choose the basis for Bessel beams. For instance, Wang et al. proposed the basis consisting solely of the TE and TM beams.[18] We have mentioned this possibility at the end of Section 3.1, but let us discuss in more details below. First, we express the TEL and TML beams of order $n$, using Eq. (63):

$$\mathbf{E}_{TEL,n}^{(x)} = \left(\mathbf{E}_{TE,n-1} + \mathbf{E}_{TE,n+1}\right)/2, \qquad \mathbf{E}_{TEL,n}^{(y)} = i\left(\mathbf{E}_{TE,n-1} - \mathbf{E}_{TE,n+1}\right)/2, \tag{79}$$

$$\mathbf{E}_{TML,n}^{(x)} = \left(\mathbf{E}_{TM,n-1} + \mathbf{E}_{TM,n+1}\right)/2, \qquad \mathbf{E}_{TML,n}^{(y)} = i\left(\mathbf{E}_{TM,n-1} - \mathbf{E}_{TM,n+1}\right)/2 \tag{80}$$

which can then be combined into any basic transverse potentials:

$$\mathbf{E}_{LE}^{(x),(y)} = k_t^{-1}\left(\pm k\mathbf{E}_{TML}^{(y),(x)} - k_z\mathbf{E}_{TEL}^{(x),(y)}\right), \tag{81}$$

$$\mathbf{E}_{LM}^{(x),(y)} = k_t^{-1}\left(\mp k\mathbf{E}_{TEL}^{(y),(x)} - k_z\mathbf{E}_{TML}^{(x),(y)}\right), \tag{82}$$

$$\mathbf{E}_{CS}^{(x),(y)} = \frac{k_t}{2(k - k_z)}\left(\pm\mathbf{E}_{TML}^{(y),(x)} - \mathbf{E}_{TEL}^{(x),(y)}\right), \tag{83}$$

$$\mathbf{E}_{CS'}^{(x),(y)} = \frac{k_t}{2(k + k_z)}\left(\mp\mathbf{E}_{TML}^{(y),(x)} - \mathbf{E}_{TEL}^{(x),(y)}\right), \tag{84}$$

where the order of all beams is the same. These expressions for the LE and CS beams are analogous to that in Ref. [18] up to constant factors due to different definitions of $f_n$. Moreover, they show that the TEL and TML beams also form a convenient basis, as well as the CS and CS′ beams. Further discussion of these bases with respect to orthogonality is postponed till Section 3.5. Interestingly, Eqs. (79), (80) demonstrate that $\mathbf{E}_{TEL}^{(\alpha,\beta)}$ and $\mathbf{H}_{TML}^{(\alpha,\beta)}$ are fully transverse (have zero $z$-component), inheriting this property from the TE and TM beams, respectively.

Finally, many recurrent relations can be derived from Eq. (44). For instance, starting with $\mathbf{E}_{LE,n}^{(1,-i)}$ or $\mathbf{E}_{LM,n}^{(1,-i)}$, the last term in Eq. (44) vanish and we obtain, respectively:

$$k_t^2\mathbf{E}_{LE,n}^{(1,-i)} = 2ikk_z\mathbf{E}_{LM,n-2}^{(1,i)} - (k^2 + k_z^2)\mathbf{E}_{LE,n-2}^{(1,i)}, \tag{85}$$

$$k_t^2\mathbf{E}_{LM,n}^{(1,-i)} = -2ikk_z\mathbf{E}_{LE,n-2}^{(1,i)} - (k^2 + k_z^2)\mathbf{E}_{LM,n-2}^{(1,i)}. \tag{86}$$

## 3.4. Quadratic functionals of fields

In this and the next section we will consider complex conjugation of fields. Thus, to simplify discussion we require the angle $\alpha_0$ to be real, which implies that both $\hat{k}_t \stackrel{\text{def}}{=} k_t/k$ and $\hat{k}_z \stackrel{\text{def}}{=} k_z/k$ are also real. We have introduced $\hat{k}_t$ and $\hat{k}_z$ to shorten further expressions, although they are equal to $\sin\alpha_0$ and $\cos\alpha_0$, respectively. Let us first rewrite some of the previous formulae in the matrix form. In particular, Eqs. (15), (17), (18) can be rewritten as

$$\nabla\times\begin{pmatrix}\mathbf{e}_+\\ \mathbf{e}_-\\ \mathbf{e}_z\end{pmatrix}f_n = k\widetilde{\mathbf{T}}'\begin{pmatrix}\mathbf{e}_+\\ \mathbf{e}_-\\ \mathbf{e}_z\end{pmatrix}f_{n'}, \quad \widetilde{\mathbf{T}}' \stackrel{\text{def}}{=} \begin{pmatrix}\hat{k}_z & 0 & -i\hat{k}_t\mathcal{N}_1\\ 0 & -\hat{k}_z & -i\hat{k}_t\mathcal{N}_{-1}\\ \frac{i\hat{k}_t}{2}\mathcal{N}_{-1} & \frac{i\hat{k}_t}{2}\mathcal{N}_1 & 0\end{pmatrix}, \tag{87}$$

where, as before the tilde and prime denotes that the matrix describes the action on full 3D vectors (not only transverse ones) and the circular transverse basis is used, respectively. If the standard Descartes basis is used, Eq. (87) is transformed into

$$\nabla\times\begin{pmatrix}\mathbf{e}_x\\ \mathbf{e}_y\\ \mathbf{e}_z\end{pmatrix}f_n = \widetilde{\mathbf{T}}\begin{pmatrix}\mathbf{e}_x\\ \mathbf{e}_y\\ \mathbf{e}_z\end{pmatrix}f_{n'}, \quad \widetilde{\mathbf{T}} \stackrel{\text{def}}{=} \widetilde{\mathbf{W}}^{-1}\widetilde{\mathbf{T}}'\widetilde{\mathbf{W}}, \tag{88}$$

where $\widetilde{\mathbf{W}}$ is the matrix $\mathbf{W}$, given by Eq. (33), extended to the size 3×3 by addition value of 1 at the position (3,3). The matrix $\widetilde{\mathbf{T}}$ has a bit longer expression, but it has zero diagonal and its Hermitian adjoint $\widetilde{\mathbf{T}}^H$ coincides with $\widetilde{\mathbf{T}}$ if the signs of order increments of operators $\mathcal{N}$ are reversed, i.e. $\mathcal{N}_1$ and $\mathcal{N}_{-1}$ are interchanged. The corresponding symmetry for the matrix $\widetilde{\mathbf{T}}'$ involves additional factors of 2 due to non-normalized basis (see Eq. (94) below).

It is easy to check that

$$\widetilde{\mathbf{T}}^3 = \widetilde{\mathbf{T}}, \quad \widetilde{\mathbf{T}}'^3 = \widetilde{\mathbf{T}}' \tag{89}$$

as an implication of the Helmholtz equation (here superscripts denote matrix powers). However, $\widetilde{\mathbf{T}}^2$ or $\widetilde{\mathbf{T}}'^2$, required for expressions of double curls, require an explicit evaluation. For instance,

$$\widetilde{\mathbf{T}}'^2 = \frac{1}{2}\begin{pmatrix}1+\hat{k}_z^2 & \hat{k}_t^2\mathcal{N}_2 & -2i\hat{k}_z\hat{k}_t\mathcal{N}_1\\ \hat{k}_t^2\mathcal{N}_{-2} & 1+\hat{k}_z^2 & 2i\hat{k}_z\hat{k}_t\mathcal{N}_{-1}\\ i\hat{k}_z\hat{k}_t\mathcal{N}_{-1} & -i\hat{k}_z\hat{k}_t\mathcal{N}_1 & 2\hat{k}_t^2\end{pmatrix}. \tag{90}$$

These transformation matrices lead to concise expressions for electric and magnetic fields generated by arbitrary Hertz vector potentials (both transverse and longitudinal). However, they are redundant, which is discussed in details in Section 3.1, but can also be described in terms of nontrivial null space of matrices $\widetilde{\mathbf{T}}$ and $\widetilde{\mathbf{T}}'$. Thus, to leave only transverse potentials (and to use matrices $\mathbf{M}$ and $\mathbf{M}'$) we truncate the matrices $\widetilde{\mathbf{T}}, \widetilde{\mathbf{T}}', \widetilde{\mathbf{T}}^2$, and $\widetilde{\mathbf{T}}'^2$, leaving only the first two rows. The resulting 2×3 matrices are denoted $\mathbf{T}, \mathbf{T}', \mathbf{T}_2$, and $\mathbf{T}_2'$, respectively. Note that subscript 2 is used instead of power index, since $\mathbf{T}_2 \neq \mathbf{T}^2$. Moreover, the relation between these matrices are similar to Eq. (88):

$$\mathbf{T} = \mathbf{W}^{-1}\mathbf{T}'\widetilde{\mathbf{W}}, \quad \mathbf{T}_2 = \mathbf{W}^{-1}\mathbf{T}_2'\widetilde{\mathbf{W}}. \tag{91}$$

Finally, Eqs. (28), (29), and (31) can be rewritten as

$$\begin{pmatrix} \mathbf{E} \\ \eta \mathbf{H} \end{pmatrix} = E_0 \mathbf{M}(\mathbf{T}_2 - i\mathbf{RT}) \begin{pmatrix} \mathbf{e}_x \\ \mathbf{e}_y \\ \mathbf{e}_z \end{pmatrix} f_n = E_0 \mathbf{M}'(\mathbf{T}'_2 - i\mathbf{RT}') \begin{pmatrix} \mathbf{e}_+ \\ \mathbf{e}_- \\ \mathbf{e}_z \end{pmatrix} f_n, \qquad (92)$$

which is equivalent to Eqs. (35), (36) and not necessarily simpler, since it requires a lot of additional definitions. The true power of this representation, however, manifests itself in computing the quadratic or sesquilinear combinations of the fields. For instance, a 2×2 matrix of vector norms and scalar products $\mathbf{E} \cdot \mathbf{H}^*$ is given as

$$\begin{pmatrix} \mathbf{E} \\ \eta \mathbf{H} \end{pmatrix} \cdot \begin{pmatrix} \mathbf{E} \\ \eta \mathbf{H} \end{pmatrix}^H = |E_0^2|(\mathbf{M}'\mathbf{T}'_2 - i\mathbf{RM}'\mathbf{T}') f_n \mathbf{C}_1 (\mathbf{M}'\mathbf{T}'_2 - i\mathbf{RM}'\mathbf{T}')^H f_n^*, \qquad (93)$$

where the matrices $\mathbf{T}'$ and $\mathbf{T}'_2$ are multiplied as usual with the only caveat that operators $\mathcal{N}_l$ in the original and Hermitian transposed matrices act on $f_n$ and $f_n^*$, respectively. The diagonal matrix $\mathbf{C}_1$ is defined as

$$\mathbf{C}_1 \stackrel{\text{def}}{=} \begin{pmatrix} \mathbf{e}_+ \\ \mathbf{e}_- \\ \mathbf{e}_z \end{pmatrix} \cdot \begin{pmatrix} \mathbf{e}_+ \\ \mathbf{e}_- \\ \mathbf{e}_z \end{pmatrix}^H = \widetilde{\mathbf{W}} \widetilde{\mathbf{W}}^H = \begin{pmatrix} 2 & 0 & 0 \\ 0 & 2 & 0 \\ 0 & 0 & 1 \end{pmatrix}. \qquad (94)$$

The first element of the resulting matrix in Eq. (93) is $|\mathbf{E}|^2$, but the explicit expression is cumbersome and can also be obtained from Eq. (35):

$$\begin{aligned} |\mathbf{E}|^2 = |E_0^2| \Big[ & \left( \left| (1 + \hat{k}_z^2) M_{\text{e},+} + 2i\hat{k}_z M_{\text{m},+} \right|^2 + \left| (1 + \hat{k}_z^2) M_{\text{e},-} - 2i\hat{k}_z M_{\text{m},-} \right|^2 \right) |f_n|^2/2 \\ & + \left( \left| M_{\text{e},-} f_{n-2} \right|^2 + \left| M_{\text{e},+} f_{n+2} \right|^2 \right) \hat{k}_t^4 / 2 \\ & + \hat{k}_t^2 \left( \left| \hat{k}_z M_{\text{e},-} - i M_{\text{m},-} \right|^2 |f_{n-1}|^2 + \left| \hat{k}_z M_{\text{e},+} + i M_{\text{m},+} \right|^2 |f_{n+1}|^2 \right) \\ & + \text{Re}\big( \hat{k}_t^2 \{ M_{\text{e},+} \big[ (1 + \hat{k}_z^2) M_{\text{e},-}^* + 2i\hat{k}_z M_{\text{m},-}^* \big] f_{n+2} f_n^* \\ & + M_{\text{e},-} \big[ (1 + \hat{k}_z^2) M_{\text{e},+}^* - 2i\hat{k}_z M_{\text{m},+}^* \big] f_{n-2} f_n^* \\ & - 2 \big( \hat{k}_z M_{\text{e},+} + i M_{\text{m},+} \big) \big( \hat{k}_z M_{\text{e},-}^* + i M_{\text{m},-}^* \big) f_{n+1} f_{n-1}^* \} \big) \Big]. \end{aligned} \qquad (95)$$

Let us instead consider a symmetric combination $w \stackrel{\text{def}}{=} |\mathbf{E}|^2 + |\eta \mathbf{H}|^2$, which is proportional to the time-averaged energy density of the electromagnetic field if the host medium is non-absorbing (i.e. both $\varepsilon$ and $\mu$ are real). The latter is conventionally given as $(\varepsilon|\mathbf{E}|^2 + \mu|\mathbf{H}|^2)/4$,[52] although other definitions have been proposed in the case of magnetic medium.[53] In a weakly absorbing medium, $w$ is approximately equal to the energy density. Using the property of a trace of matrix product, i.e. $\text{tr}(\mathbf{AB}) = \text{tr}(\mathbf{BA})$, we obtain

$$w = |E_0^2| \, \text{tr}[\mathbf{X}_1 (\mathbf{M}'^H \mathbf{M}') - \mathbf{Y}_1 (\mathbf{M}'^H i \mathbf{R} \mathbf{M}')], \qquad (96)$$

where the Hermitian matrices $\mathbf{X}_1$ and $\mathbf{Y}_1$ are defined as:

$$\mathbf{X}_1 \stackrel{\text{def}}{=} \mathbf{T}' f_n \mathbf{C}_1 \mathbf{T}'^H f_n^* + \mathbf{T}'_2 f_n \mathbf{C}_1 \mathbf{T}'^H_2 f_n^* = \frac{1}{2}\Big[(\hat{k}_t^4 + 8\hat{k}_z^2)|f_n|^2 \begin{pmatrix} 1 & 0 \\ 0 & 1 \end{pmatrix}$$
$$+ \hat{k}_t^4 \begin{pmatrix} |f_{n+2}|^2 & 2f_{n+1}f_{n-1}^* \\ \text{c.c.} & |f_{n-2}|^2 \end{pmatrix} + (1 - \hat{k}_z^4)\begin{pmatrix} 2|f_{n+1}|^2 & f_n f_{n-2}^* + f_{n+2}f_n^* \\ \text{c.c.} & 2|f_{n-1}|^2 \end{pmatrix}\Big], \tag{97}$$

$$\mathbf{Y}_1 \stackrel{\text{def}}{=} \mathbf{T}' f_n \mathbf{C}_1 \mathbf{T}'^H_2 f_n^* + \mathbf{T}'_2 f_n \mathbf{C}_1 \mathbf{T}'^H f_n^*$$
$$= 2\hat{k}_z(1 + \hat{k}_z^2)|f_n|^2 \begin{pmatrix} 1 & 0 \\ 0 & -1 \end{pmatrix} + \hat{k}_t^2 \hat{k}_z \begin{pmatrix} 2|f_{n+1}|^2 & f_n f_{n-2}^* - f_{n+2}f_n^* \\ \text{c.c.} & -2|f_{n-1}|^2 \end{pmatrix}, \tag{98}$$

where "c.c." denotes complex conjugate of the another off-diagonal term in the same matrix.

Eq. (96) consists of traces of products of two Hermitian matrices, each equals the element-wise inner product of these matrices. Thus, albeit being still cumbersome $w$ has a simple structure of linear combinations of various products of $f_n$ with weights given by elements of matrices $\mathbf{M}'^H \mathbf{M}'$ and $\mathbf{M}'^H \mathbf{iRM}'$. Obviously, Eq. (12) implies

$$f_n f_l^* = J_n(k_t \rho) J_l^*(k_t \rho) e^{i(n-l)\varphi} e^{-2z \operatorname{Im} k_z}, \tag{99}$$

which means that the diagonals of matrices $\mathbf{X}_1$ and $\mathbf{Y}_1$ contribute to $\varphi$-independent terms, while off-diagonal terms – to $\varphi$-dependent ones.

The first important conclusion is that $w$ is circularly symmetric (independent of $\varphi$) if and only if both $\mathbf{M}'^H \mathbf{M}'$ and $\mathbf{M}'^H \mathbf{RM}'$ are diagonal, since the terms $f_{n+1} f_{n-1}^*$ in $\mathbf{X}_1$ cannot be compensated by any terms in $\mathbf{Y}_1$. This is equivalent to the condition

$$M_{e,+} M_{e,-}^* + M_{m,+} M_{m,-}^* = 0 = M_{e,+} M_{m,-}^* - M_{m,+} M_{e,-}^*, \tag{100}$$

which, in turn, is equivalent to either 1) $\mathbf{M}'_+ = \mathbf{0}$ or $\mathbf{M}'_- = \mathbf{0}$, i.e. any linear combination of the TE and TM beams of the same order [cf. Eqs. (46), (72)] or 2) all elements of $\mathbf{M}'$ are not zero and

$$\frac{M_{e,+}}{M_{m,+}} = \frac{M_{m,-}}{M_{e,-}} = \pm i. \tag{101}$$

In terms of matrix $\mathbf{M}$, Eq. (101) is equivalent to

$$M_{e,x} = \pm M_{m,y} \quad \text{and} \quad M_{e,y} = \mp M_{m,x}, \tag{102}$$

where $\pm$ sign corresponds to that in Eq. (101). Taking a look at Table 2 one can see that the corresponding matrix is either $\mathbf{M}_{CS}^{(\alpha,\beta)}$ or $\mathbf{M}_{CS'}^{(\alpha,\beta)}$. On the one hand, we have justified the name CS for these beams – they indeed have circularly symmetric $w$ (see Figure 7 and Figure 11). On the other hand, we have enumerated all Bessel beams described by a single matrix $\mathbf{M}$, possessing such symmetry.

One-way implication, i.e. the symmetry of the TE, TM and CS beams is known[17,18] and can be easily derived from their properties (as well as for the CS′ beam). The rotation of the field is given by Eq. (51), where the phase factor vanishes after multiplication by its transpose in sesquilinear combinations of fields. Thus, Eq. (54) implies circular symmetry of both $|\mathbf{E}|^2$, $|\mathbf{H}|^2$, and, hence, of $w$ for the TE and TM beams. For the CS and CS′ beams, Eqs. (73) and (74) allow one to move $\mathbf{R}_\psi$ and $\mathbf{R}_{-\psi}$ to the outer sides of the right-hand side of Eq. (93), where they cancel each other, after taking the trace, proving the circular symmetry of $w$. In other words, for the latter beams the polarization rotation is replaced by duality rotation, to which $w$ is obviously invariant.

The circular symmetry of $|\mathbf{E}|^2$ is harder to achieve – it requires all three quadratic combinations of coefficients inside $\mathrm{Re}(...)$ in Eq. (95) to vanish. This holds if and only if either $\mathbf{M}'_+ = \mathbf{0}$ or $\mathbf{M}'_- = \mathbf{0}$, i.e. for any linear combination of the TE and TM beams of the same order. Again, this symmetry is known,[18] but we also proved that no other Bessel beam described by a matrix $\mathbf{M}$ has such symmetry. Due to the duality the same requirement holds for the circular symmetry of $|\mathbf{H}|^2$.

We are further interested in the time-averaged Poynting vector[52]

$$\mathbf{S} = \frac{1}{2}\mathrm{Re}(\mathbf{E}\times\mathbf{H}^*), \quad (103)$$

although three different alternative expressions exist.[54,55] Analogously to Eq. (93) let us consider the 2×2 skew-Hermitian matrix

$$\begin{pmatrix}\mathbf{E}\\\eta\mathbf{H}\end{pmatrix}\times\begin{pmatrix}\mathbf{E}\\\eta\mathbf{H}\end{pmatrix}^H = |E_0^2|(\mathbf{M}'\mathbf{T}'_2 - \mathrm{i}\mathbf{R}\mathbf{M}'\mathbf{T}')f_n\mathrm{i}\mathbf{C}_2(\mathbf{M}'\mathbf{T}'_2 - \mathrm{i}\mathbf{R}\mathbf{M}'\mathbf{T}')^H f_n^*, \quad (104)$$

where $\mathbf{C}_2$ is the Hermitian matrix of vectors, obtained using Eq. (13):

$$\mathbf{C}_2 \stackrel{\mathrm{def}}{=} -\mathrm{i}\begin{pmatrix}\mathbf{e}_+\\\mathbf{e}_-\\\mathbf{e}_z\end{pmatrix}\times\begin{pmatrix}\mathbf{e}_+\\\mathbf{e}_-\\\mathbf{e}_z\end{pmatrix}^H = \begin{pmatrix}-2\mathbf{e}_z & 0 & \mathbf{e}_+\\ 0 & 2\mathbf{e}_z & -\mathbf{e}_-\\ \mathbf{e}_- & -\mathbf{e}_+ & 0\end{pmatrix}. \quad (105)$$

We will further calculate the expression $2\,\mathrm{Re}[\mathbf{E}\times(\eta\mathbf{H})^*]$ by combining off-diagonal elements in Eq. (104). This expression is proportional both to Eq. (103) and to alternative definitions in a non-absorbing host medium, and is approximately proportional to them in a weakly absorbing one, similar to the discussion of the energy density above.

Taking the trace of $\mathbf{R}$ times Eq. (104), we obtain

$$2\,\mathrm{Re}[\mathbf{E}\times(\eta\mathbf{H})^*] = |E_0^2|\,\mathrm{tr}[\mathbf{X}_2(\mathbf{M}'^H\mathrm{i}\mathbf{R}\mathbf{M}') - \mathbf{Y}_2(\mathbf{M}'^H\mathbf{M}')], \quad (106)$$

where the Hermitian matrices $\mathbf{X}_2$ and $\mathbf{Y}_2$ are defined analogously to Eqs. (97), (98), but their elements are vector functions:

$$\mathbf{X}_2 \stackrel{\mathrm{def}}{=} \mathbf{T}'f_n\mathbf{C}_2\mathbf{T}'^H f_n^* + \mathbf{T}'_2 f_n\mathbf{C}_2\mathbf{T}'^H_2 f_n^* = \frac{\mathbf{e}_z}{2}\left[(\hat{k}_t^4 + 8\hat{k}_z^2)|f_n|^2\begin{pmatrix}-1 & 0\\0 & 1\end{pmatrix}\right.$$
$$+ \hat{k}_t^2\begin{pmatrix}\hat{k}_t^2|f_{n+2}|^2 & (1+\hat{k}_z^2)(f_{n+2}f_n^* - f_n f_{n-2}^*)\\ \mathrm{c.\,c.} & -\hat{k}_t^2|f_{n-2}|^2\end{pmatrix}\right] \quad (107)$$
$$+ \frac{\hat{k}_z\hat{k}_t}{2}\left[(3+\hat{k}_z^2)\begin{pmatrix}-\mathbf{g}_n & 0\\ 0 & \mathbf{g}_{n-1}\end{pmatrix} + \hat{k}_t^2\begin{pmatrix}-\mathbf{g}_{n+1} & \mathbf{h}_{n+1}-\mathbf{h}_n\\ \mathrm{c.\,c.} & \mathbf{g}_{n-2}\end{pmatrix}\right],$$

$$\mathbf{Y}_2 \stackrel{\text{def}}{=} \mathbf{T}'f_n\mathbf{C}_2\mathbf{T}_2'^H f_n^* + \mathbf{T}_2'f_n\mathbf{C}_2\mathbf{T}'^H f_n^*$$

$$= -\hat{k}_z\mathbf{e}_z\begin{pmatrix} 2(1+\hat{k}_z^2)|f_n|^2 & \hat{k}_t^2(f_{n+2}f_n^* + f_nf_{n-2}^*) \\ \text{c.c.} & 2(1+\hat{k}_z^2)|f_n|^2 \end{pmatrix} \tag{108}$$

$$-\frac{\hat{k}_t}{2}\left[(1+3\hat{k}_z^2)\begin{pmatrix} \mathbf{g}_n & 0 \\ 0 & \mathbf{g}_{n-1}\end{pmatrix} + \hat{k}_t^2\begin{pmatrix} \mathbf{g}_{n+1} & \mathbf{h}_n + \mathbf{h}_{n+1} \\ \text{c.c.} & \mathbf{g}_{n-2}\end{pmatrix}\right],$$

and the transverse vector functions $\mathbf{g}_n$ and $\mathbf{h}_n$ are the following:

$$\mathbf{g}_n \stackrel{\text{def}}{=} \mathrm{i}(\mathbf{e}_-f_{n+1}f_n^* - \mathbf{e}_+f_nf_{n+1}^*) = 2\,\mathrm{Im}(\mathbf{e}_+f_nf_{n+1}^*), \tag{109}$$

$$\mathbf{h}_n \stackrel{\text{def}}{=} \mathrm{i}(\mathbf{e}_-f_{n+1}f_{n-2}^* - \mathbf{e}_+f_nf_{n-1}^*). \tag{110}$$

Again, the final expressions are cumbersome but allow straightforward analysis. First, the z-component of Eq. (106) (in many cases proportional to $S_z$) is circularly symmetric if and only if both $\mathbf{M}'^H\mathbf{M}'$ and $\mathbf{M}'^H\mathbf{R}\mathbf{M}'$ are diagonal (completely analogous to $w$). In this case the transverse component of Eq. (106) is a linear combination of $\mathbf{g}_n$. On the one hand, this combination never vanishes (unless $\alpha \to 0$) due to slightly different coefficients in Eqs. (107) and (108) – $(3k^2 + k_z^2)$ versus $(k^2 + 3k_z^2)$. On the other hand, $\mathbf{g}_n$ has a simple azimuthal dependence:

$$\mathbf{g}_n = 2\,\mathrm{Im}\big[(\mathbf{e}_\rho + \mathrm{i}\mathbf{e}_\varphi)J_n(k_t\rho)J_l^*(k_t\rho)\big]e^{-2z\,\mathrm{Im}\,k_z}, \tag{111}$$

i.e. it rotates with $\varphi$ the same way as the unit vectors of the cylindrical coordinate system $\mathbf{e}_\rho$ and $\mathbf{e}_\varphi$. Moreover, for real $k$ the vector $\mathbf{g}_n$ is always directed along $\mathbf{e}_\varphi$. But even for complex $k$, any linear combination of $\mathbf{g}_n$ would not change its magnitude with $\varphi$. Thus, we have proved that for Bessel beams, given by Eq. (31) in non-absorbing host medium (for real $k$), both $S_z$ and $|\mathbf{S}_t|$ (and, hence, $|\mathbf{S}|$) are circularly symmetric if and only if $w$ has the same property, i.e. the beam is either CS, CS′, or a linear combination of the TE and TM beams. This symmetry is well known for the TE, TM and CS[17,18] and can be derived directly similarly to the case of $w$ above, but here we present a complete list of conforming beams. For weakly absorbing medium the above symmetry property is approximately satisfied.

To finalize this section let us show an example of the CS′ beam profiles in Figure 10 and Figure 11, since it has not been previously discussed in the literature.

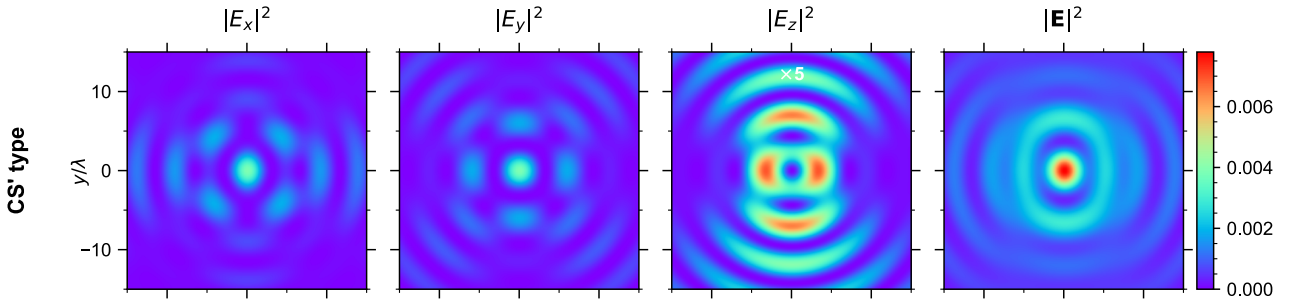

**Figure 10.** Intensity profiles of components of $\mathbf{E}_{\text{CS}'}^{(x)}$ with the same parameters as in Figure 2. The z-component is scaled for better visibility.

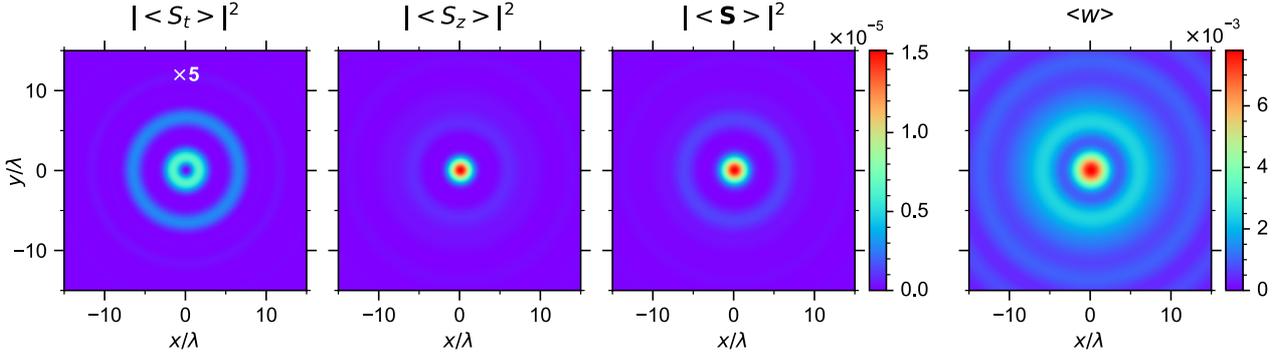

**Figure 11.** Magnitude profiles of the Poynting vector components and energy density for $\mathbf{E}_{CS}^{(x)}$, with the same parameters as in Figure 2. The transverse component is scaled for better visibility.

### 3.5. Orthogonality and norm

The discussion of relations of different Bessel beam types naturally leads to a question: what is the basis of Bessel beams? While there is no unique answer to this question, one can at least examine the orthogonality and norm relations. Any reasonable norm for electromagnetic fields is based on the quadratic functionals, described in the previous section, such as $w$, $|\mathbf{E}|^2$, and $|\mathbf{H}|^2$. However, they should be integrated over some domains to become independent of $\mathbf{r}$. Since the idealized Bessel beams have trivial dependence on $z$, integration over this coordinate is meaningless in contrast to the transverse coordinate $\boldsymbol{\rho}$. However, plain integration over $\boldsymbol{\rho}$ leads to infinite result due to slow decay of the Bessel functions, similar to the case of plane waves. Hence, some renormalization is required. Using asymptotic expansion of a Bessel function for large (complex) arguments (Eq. 10.17.3 of Ref. [56]) one can obtain

$$\int_{\rho \leq R} d^2\boldsymbol{\rho}\, |J_n(k_t\rho)|^2 = 2\pi \int_0^R d\rho\, \rho |J_n(k_t\rho)|^2 \sim \xi^{-1}(k_t, R), \qquad (112)$$

where the normalization function is

$$\xi(k_t, R) \stackrel{\text{def}}{=} \frac{|k_t|\, \text{Im}\, k_t}{\sinh(2R\, \text{Im}\, k_t)} \xrightarrow[\text{Im}\, k_t \to 0]{} \frac{|k_t|}{2R}. \qquad (113)$$

Thus, we further assume $z = 0$ in this section and define the finite norm of electric field of Bessel beams as

$$\|\mathbf{E}(\mathbf{r})\|^2 \stackrel{\text{def}}{=} \lim_{R \to \infty} \xi(k_t, R) \int_{\rho \leq R} d^2\boldsymbol{\rho}\, |\mathbf{E}(\mathbf{r})|^2, \qquad (114)$$

which can also be applied to the magnetic field. Note that in the weakly absorbing medium (i.e. when $\text{Im}\, k_t \ll |k_t|$) the limiting value is obtained already for moderate values of $R$, when the limit of $\xi(k_t, R)$ in Eq. (113) is approximately valid.

A natural next step is to define the functional inner product conforming to the above norm:

$$\langle \mathbf{E}_1, \mathbf{E}_2 \rangle \overset{\text{def}}{=} \lim_{R \to \infty} \xi(k_t, R) \int_{\rho \leq R} d^2\boldsymbol{\rho} \, \mathbf{E}_1(\mathbf{r}) \cdot \mathbf{E}_2^*(\mathbf{r}), \tag{115}$$

which we further evaluate for two beams specified by arbitrary matrices $\mathbf{M}_1$ and $\mathbf{M}_2$. For that we use the following equality (for $z = 0$), which follows from Eqs. (99) and (112):

$$\lim_{R \to \infty} \xi(k_t, R) \int_{\rho \leq R} d^2\boldsymbol{\rho} \, f_n f_l^* = \delta_{nl}, \tag{116}$$

where $\delta_{nl}$ is the Kronecker symbol. Let us start from Eq. (93) generalized to two different beams and integrate it over the beam cross section:

$$\left\langle \begin{pmatrix} \mathbf{E}_1 \\ \eta \mathbf{H}_1 \end{pmatrix}, \begin{pmatrix} \mathbf{E}_2 \\ \eta \mathbf{H}_2 \end{pmatrix} \right\rangle$$
$$= |E_0^2| \left[ (1 + \hat{k}_z^2)(\mathbf{M}_1' \mathbf{M}_2'^H - \mathbf{R} \mathbf{M}_1' \mathbf{M}_2'^H \mathbf{R}) - 2i\hat{k}_z (\mathbf{R} \mathbf{M}_1' \mathbf{I}' \mathbf{M}_2'^H + \mathbf{M}_1' \mathbf{I}' \mathbf{M}_2'^H \mathbf{R}) \right], \tag{117}$$

where $\mathbf{I}'$ is the reflection matrix

$$\mathbf{I}' \overset{\text{def}}{=} \begin{pmatrix} 1 & 0 \\ 0 & -1 \end{pmatrix} \tag{118}$$

and we used the equivalence with respect to calculation of integrals, following from Eq. (116):

$$\mathbf{T}' \mathbf{C}_1 \mathbf{T}'^H \sim \mathbf{T}_2' \mathbf{C}_1 \mathbf{T}_2'^H \sim (1 + \hat{k}_z^2)\mathbf{I}, \quad \mathbf{T}' \mathbf{C}_1 \mathbf{T}_2'^H \sim \mathbf{T}_2' \mathbf{C}_1 \mathbf{T}'^H \sim 2\hat{k}_z \mathbf{I}' \tag{119}$$

[cf. also Eq. (97), (98)].

Equation (117) implies that both diagonal elements of the resulting matrix are equal and can be concisely expressed through its trace:

$$\langle \mathbf{E}_1, \mathbf{E}_2 \rangle = |\eta|^2 \langle \mathbf{H}_1, \mathbf{H}_2 \rangle = |E_0^2| \operatorname{tr}\left[ (1 + \hat{k}_z^2) \mathbf{M}_1' \mathbf{M}_2'^H - 2i\hat{k}_z \mathbf{R} \mathbf{M}_1' \mathbf{I}' \mathbf{M}_2'^H \right]$$
$$= |E_0^2| \operatorname{tr}\left[ (1 + \hat{k}_z^2) \mathbf{M}_1 \mathbf{M}_2^H - 2\hat{k}_z \mathbf{R} \mathbf{M}_1 \mathbf{R} \mathbf{M}_2^H \right]/2, \tag{120}$$

where the last part follows from Eq. (32) and identities $\mathbf{W}^{-1} \mathbf{W}^{-H} = \mathbf{I}/2$, $\mathbf{W}^{-1} \mathbf{I}' \mathbf{W}^{-H} = -i\mathbf{R}/2$. One can also define the norm of the beam through $w$ – the conforming inner product will be equal to $2\langle \mathbf{E}_1, \mathbf{E}_2 \rangle$. As expected, this inner product is invariant to both duality and polarization rotation, which is evident from the corresponding transformations of matrices $\mathbf{M}_1$ and $\mathbf{M}_2$ (see Section 3.1). Moreover, beams with two orthogonal polarizations, i.e. with matrices $\mathbf{M}$ and $\mathbf{MR}$ [Eq. (53)], have zero inner product whenever

$$\operatorname{tr}(\mathbf{R}\mathbf{M}^H \mathbf{M}) = \operatorname{tr}(\mathbf{R}\mathbf{M}\mathbf{M}^H) = 0. \tag{121}$$

This is true for arbitrary real matrix $\mathbf{M}$ (due to the trace of a product of a symmetric and a skew-symmetric matrix being zero), as well as for any its linear transformation of the form $\alpha_1 \mathbf{M} + \beta_1 \mathbf{RMR}$ ($\forall \alpha_1, \beta_1 \in \mathbb{C}$). This obviously includes the LE, LM, CS, CS′, TEL and TML beams with any "real" polarization (given by $\alpha, \beta \in \mathbb{R}$, see Table 2). Based on this rigorous analysis, we define the linear and elliptical polarizations of the Bessel beams as the ones, for which Eq. (121) is valid and not valid, respectively. Constructing the second (orthogonal) polarization and using $(\alpha, \beta)$ notation should be done only when starting with a linear polarization. In particular, the normalization, given by Eq. (68), then preserves the beam norm. And any $(\alpha, \beta)$ combination with $\alpha/\beta \notin \mathbb{R}$, applied to a linear polarization, necessarily leads to an elliptical (or circular) polarization. The circular polarization can generally be defined as an eigenfunction of the rotation operator [Eq. (70)], which includes the TE and TM beams.

It is straightforward to compute the norms of various beams using Eq. (120) (and $\|\mathbf{E}(\mathbf{r})\|^2 = \langle \mathbf{E}, \mathbf{E} \rangle$) – the results are summarized in Table 4. Different norms of the CS and CS′ beams correspond to their plane-wave limits (see Table 1), however this relation is not that trivial. The plane-wave limit is taken for fixed $\mathbf{r}$, i.e. effectively for $\mathbf{r} = \mathbf{0}$, while the norm is computed over the whole cross sections, including the maxima of the fields located at large values of $\rho$ (at certain fixed values of $k_t\rho$). This difference is crucial for beams that have zero field intensity in the center.

**Table 4.** Norms of various Bessel beam types. The results for $y$-polarizations are the same as shown for $x$-ones. Results for $(\alpha, \beta)$ polarizations differ by a factor of $\sqrt{|\alpha|^2 + |\beta|^2}$.

| Type | $\|\mathbf{E}\|/E_0$ |
|---|---|
| $\mathbf{E}_{\text{LE}}^{(x)}, \mathbf{E}_{\text{LM}}^{(x)}$ | $\sqrt{(1 + \hat{k}_z^2)/2}$ |
| $\mathbf{E}_{\text{CS}}^{(x)}$ | $(1 + \hat{k}_z)/2$ |
| $\mathbf{E}_{\text{CS'}}^{(x)}$ | $(1 - \hat{k}_z)/2$ |
| $\mathbf{E}_{\text{TEL}}^{(x)}, \mathbf{E}_{\text{TML}}^{(x)}$ | $\hat{k}_t/\sqrt{2}$ |
| $\mathbf{E}_{\text{TE}}, \mathbf{E}_{\text{TM}}$ | $\hat{k}_t$ |

Let us further discuss the cross products of the Bessel beams. While the LE and LM beams have the simplest matrices $\mathbf{M}$ (see Table 3), they do not constitute an orthogonal basis due to the last term in Eq. (120). In particular,

$$\langle \mathbf{E}_{\text{LM}}^{(x)}, \mathbf{E}_{\text{LE}}^{(y)} \rangle = -\langle \mathbf{E}_{\text{LE}}^{(x)}, \mathbf{E}_{\text{LM}}^{(y)} \rangle = \hat{k}_z |E_0^2|, \tag{122}$$

while other cross products are zero due to the factors being connected by duality or polarization rotations. Alternatively, a beam with diagonal $\mathbf{M}$ is always orthogonal to the beam with zero-diagonal $\mathbf{M}$. The same arguments apply to the sets $\{\mathbf{E}_{\text{CS}}^{(x),(y)}, \mathbf{E}_{\text{CS'}}^{(x),(y)}\}$ and $\{\mathbf{E}_{\text{TEL}}^{(x),(y)}, \mathbf{E}_{\text{TML}}^{(x),(y)}\}$ proving the vanishing of most cross products. One only need to check manually the inner products involving diagonal matrices, which happen to vanish as well:

$$\langle \mathbf{E}_{\text{CS}}^{(x)}, \mathbf{E}_{\text{CS'}}^{(x)} \rangle = \langle \mathbf{E}_{\text{TEL}}^{(x)}, \mathbf{E}_{\text{TML}}^{(y)} \rangle = 0. \tag{123}$$

Thus, we have identified two orthogonal bases for beams of a fixed order specified by any matrix **M**: {CS,CS′} and {TEL,TML} (using two orthogonal linear polarizations for each). They can easily be made orthonormal dividing by the corresponding norms (Table 4). After normalization the pairs of these basic vectors are related by rotation, e.g., those in Eq. (123) , [cf. Eqs. (83), (84)]. Obviously any other (four-dimensional) rotation of a set of four orthonormal beams will produce another orthonormal basis. For instance, another obvious option is $\{\mathbf{E}_{\text{TEL}}^{(1,\pm i)}, \mathbf{E}_{\text{TML}}^{(1,\pm i)}\}$, which corresponds to the TE and TM beams of orders $n \pm 1$ [Eq. (63)] or, more generally, to orthogonal basis of CVWFs (see Sections 2.1 and 3.1). Interestingly, the basis of {CS, CS′} is closely related to the set of Pauli matrices.

Finally, let us discuss the orthogonality of beams of various orders, more specifically, with orders $n$ and $n + l$ (for fixed $l > 0$). Let us start from the known relations for the TE and TM beams. Orthogonality of these beams for $l = 2$ follows from the discussion above (they are parts of the same orthonormal basis), while for all other $l \neq 0$ the orthogonality follows immediately from Eqs. (19), (20), and (116), since the same orders of $f_n$ and $f_n^*$ appear only for $l = 1$ and only in different vector components, e.g., in transverse and longitudinal ones. This, again, is a well-known orthogonality of the set of CVWFs.[46] Together with Eqs. (79), (80) it immediately implies that two beams with any matrices $\mathbf{M}_1$ and $\mathbf{M}_2$ are always orthogonal for $l \neq 2$ [also obvious from Eq. (35)].

For the case of $l = 2$, we first consider the basis of the TEL and TML beams, for which we obtain

$$\langle \mathbf{E}_{\text{TEL},n}^{(x)}, \mathbf{E}_{\text{TEL},n+2}^{(x)} \rangle = - \langle \mathbf{E}_{\text{TEL},n}^{(y)}, \mathbf{E}_{\text{TEL},n+2}^{(y)} \rangle = \hat{k}_t^2/4,$$
$$\langle \mathbf{E}_{\text{TEL},n}^{(x)}, \mathbf{E}_{\text{TEL},n+2}^{(y)} \rangle = \langle \mathbf{E}_{\text{TEL},n}^{(y)}, \mathbf{E}_{\text{TEL},n+2}^{(x)} \rangle = -\mathrm{i}\hat{k}_t^2/4,$$
(124)

and exactly the same for TML, while the TEL and TML beams remain orthogonal to each other for any polarizations. The additional minus sign appears during rotation (switching between $x$- and $y$-polarizations) due to phase factor in Eq. (51), which is opposite for orders $n$ and $n + 2$. This can be continued to calculate cross products of other beams, but let us, first, derive a general expression for any two matrices $\mathbf{M}_1$ and $\mathbf{M}_2$ from the first principles. Analogously to Eq. (117), we obtain

$$\left\langle \begin{pmatrix} \mathbf{E}_1 \\ \eta \mathbf{H}_1 \end{pmatrix}_n, \begin{pmatrix} \mathbf{E}_2 \\ \eta \mathbf{H}_2 \end{pmatrix}_{n+2} \right\rangle = |E_0^2| \hat{k}_t^2 \left( \mathbf{M}_1' \begin{pmatrix} 0 & 1 \\ 0 & 0 \end{pmatrix} \mathbf{M}_2'^H - \mathbf{R}\mathbf{M}_1' \begin{pmatrix} 0 & 1 \\ 0 & 0 \end{pmatrix} \mathbf{M}_2'^H \mathbf{R} \right), \quad (125)$$

where we used the equivalence with respect to calculation of the integrals:

$$\mathbf{T}'\mathbf{C}_1\mathbf{T}'^H \sim \mathbf{T}_2'\mathbf{C}_1\mathbf{T}_2'^H \sim \hat{k}_t^2 \begin{pmatrix} 0 & \mathcal{N}_2 \\ 0 & 0 \end{pmatrix}, \qquad \mathbf{T}'\mathbf{C}_1\mathbf{T}_2'^H \sim \mathbf{T}_2'\mathbf{C}_1\mathbf{T}'^H \sim 0 \quad (126)$$

[cf. also Eqs. (97), (98)]. Again, both diagonal elements of the resulting matrix are equal and can be concisely expressed through its trace:

$$\langle \mathbf{E}_{1,n}, \mathbf{E}_{2,n+2} \rangle = |\eta|^2 \langle \mathbf{H}_{1,n}, \mathbf{H}_{2,n+2} \rangle$$
$$= |E_0^2|\hat{k}_t^2 \operatorname{tr}\left( \mathbf{M}_1' \begin{pmatrix} 0 & 1 \\ 0 & 0 \end{pmatrix} \mathbf{M}_2'^H \right) = |E_0^2| \frac{\hat{k}_t^2}{4} \operatorname{tr}\left( \mathbf{M}_1 \begin{pmatrix} 1 & -i \\ -i & -1 \end{pmatrix} \mathbf{M}_2^H \right). \tag{127}$$

The expressions for any specific beams, including Eq. (124) and orthogonality of the TE and TM beams, follow trivially from Eq. (127). Moreover, any beam of order $n$ with real matrix $\mathbf{M}$ is orthogonal to the dual beam (with matrix $\mathbf{RM}$) of order $n+2$, since the argument of trace is then the product of $\mathbf{R}$ and a symmetric matrix. Specifically, the CS beams are mutually orthogonal even with changing orders, the same holds for the CS′ beams. However, their cross products are not zero:

$$\langle \mathbf{E}_{\text{CS},n}^{(x)}, \mathbf{E}_{\text{CS}',n+2}^{(x)} \rangle = -\langle \mathbf{E}_{\text{CS},n}^{(y)}, \mathbf{E}_{\text{CS}',n+2}^{(y)} \rangle = \hat{k}_t^2/8,$$
$$\langle \mathbf{E}_{\text{CS},n}^{(x)}, \mathbf{E}_{\text{CS}',n+2}^{(y)} \rangle = \langle \mathbf{E}_{\text{CS},n}^{(y)}, \mathbf{E}_{\text{CS}',n+2}^{(x)} \rangle = -i\hat{k}_t^2/8, \tag{128}$$

Similarly, the LE beams are always orthogonal to the LM beams (for any two polarizations), while the cross products of the LE beams are minus that for the TEL beams [Eq. (124)]:

$$\langle \mathbf{E}_{\text{LE},n}^{(x)}, \mathbf{E}_{\text{LE},n+2}^{(x)} \rangle = -\langle \mathbf{E}_{\text{LE},n}^{(y)}, \mathbf{E}_{\text{LE},n+2}^{(y)} \rangle = -\hat{k}_t^2/4,$$
$$\langle \mathbf{E}_{\text{LE},n}^{(x)}, \mathbf{E}_{\text{LE},n+2}^{(y)} \rangle = \langle \mathbf{E}_{\text{LE},n}^{(y)}, \mathbf{E}_{\text{LE},n+2}^{(x)} \rangle = i\hat{k}_t^2/4. \tag{129}$$

The results for the LM beams are the same.

To conclude this section, we have shown that the basis based on either TEL and TML or CS and CS′ beams is orthogonal and complete, if we consider every second odd and every second even order, for instance, $n = 4l$ and $n = 4l + 1$ for any integer $l$ (see also Section 3.1). Thus, they are comparable to the well-known basis of the TE and TM beams of all orders $n$. The basis of the LE and LM beams, which directly corresponds to elements of the matrix $\mathbf{M}$, is not fully orthogonal, but is the most convenient one, when a single order $n$ is considered, due to simple rotation and duality transformations and ability to express all other beam types. Thus, we chose it for implementation in the light-scattering simulation code as discussed in the next section.

## 4. Scattering of Bessel beams

### 4.1. Extension of the Mueller calculus to Bessel beams

Most of the light scattering codes, including ADDA, are tailored for the Mueller calculus that implies consideration of two polarizations of incident electromagnetic field.[23] This calculus is typically considered for non-magnetic materials and non-absorbing host medium; the same limitation is currently present in ADDA. So we further assume real $k$ and $\mu = 1$ and discuss only the electric field. The amplitude scattering matrix, by definition, relates polarizations of incident and scattered fields, expanded into a basis of two orthogonal components, so-called Jones vectors:[23]

$$\begin{pmatrix} E_\parallel \\ E_\perp \end{pmatrix}_{\text{sca}} = \frac{e^{ik(r-z)}}{-ikr} \begin{pmatrix} S_2 & S_3 \\ S_4 & S_1 \end{pmatrix} \begin{pmatrix} E_\parallel \\ E_\perp \end{pmatrix}_{\text{inc}}, \tag{130}$$

The central role in scattering problems is played by the Stokes parameters $(I, Q, U, V)$ which are quadratic combinations of the fields, when the latter are fully coherent:[23]

$$I = E_\parallel E_\parallel^* + E_\perp E_\perp^*, \tag{131}$$

$$Q = E_\parallel E_\parallel^* - E_\perp E_\perp^*, \tag{132}$$

$$U = E_\parallel E_\perp^* + E_\perp E_\parallel^*, \tag{133}$$

$$V = i(E_\parallel E_\perp^* - E_\perp E_\parallel^*). \tag{134}$$

Originally, Stokes parameters were proposed as a set of measurable quantities that describe polarization states for plane waves[23] (e.g., $I$ is the intensity). Importantly, the Stokes vectors for the scattered and incident waves are linearly related through the Mueller (scattering) matrix

$$\begin{pmatrix} I \\ Q \\ U \\ V \end{pmatrix}_{\text{sca}} = \frac{1}{k^2 r^2} \begin{pmatrix} S_{11} & S_{12} & S_{13} & S_{14} \\ S_{21} & S_{22} & S_{23} & S_{24} \\ S_{31} & S_{32} & S_{33} & S_{34} \\ S_{41} & S_{42} & S_{43} & S_{44} \end{pmatrix} \begin{pmatrix} I \\ Q \\ U \\ V \end{pmatrix}_{\text{inc}}. \tag{135}$$

The elements of the Mueller matrix can be expressed through that of the amplitude scattering matrix[23] and allow expressing the measured signal for any detector configuration. In particular, the scattering intensities for parallel and perpendicular incident polarizations are given as:

$$I_\parallel = S_{11} + S_{12}, \tag{136}$$

$$I_\perp = S_{11} - S_{12}, \tag{137}$$

which, for axisymmetric particles, are equivalent to scattering intensities in parallel and perpendicular scattering planes (E-plane and H-plane, respectively) for the same incident polarization.

In the following, we extend the Mueller calculus to the Bessel beams. For that we focus on the linear relation (130), which remains valid for any definition of two incident polarizations (basis components). Specifically, once one solves the scattering problem for two incident fields and obtains the corresponding amplitude matrix, Eq. (130) provides the solution of the scattering problem for any linear combination of these two incident fields. In contrast to the case of a plane wave, such linear combinations do not cover the full range of incident fields. Still, if the two polarizations are related by a rotation, the linear combinations will cover all their possible rotations. Note that both Jones and Mueller vectors remain intact for the scattered field, since the latter is equivalent to a plane wave in the far field irrespective of the incident beam.

Next, we postulate that two basic fields (polarizations) of the incident beam $\mathbf{E}^\perp(\mathbf{r})$ and $\mathbf{E}^\parallel(\mathbf{r})$ are related by Eq. (53), i.e. by a $\pi/2$ rotation with additional phase factor. Omitting further the dependence on $\mathbf{r}$, this definition corresponds to $\{\mathbf{E}^{(x)}, \mathbf{E}^{(y)}\}$ for most of the Bessel beam types, as discussed in Section 3.2. The components $E_\parallel$ and $E_\perp$ then exactly correspond to $\beta$ and $\alpha$, respectively, in definition of $\mathbf{E}^{(\alpha,\beta)}$ [Eq. (62)]. The advantage of such postulate is the simplest rotation relations, based on Eq. (69). Specifically, rotating the beam by angle $\psi$ over the $z$-axis transforms its components as

$$\begin{pmatrix} E_\parallel \\ E_\perp \end{pmatrix}'_{\text{inc}} = e^{-in\psi} \mathbf{R}_{-\psi} \begin{pmatrix} E_\parallel \\ E_\perp \end{pmatrix}_{\text{inc}}. \tag{138}$$

The same rotation of the coordinate frame (by angle $\psi$), which corresponds, e.g., to rotation of the scatterer, is equivalent to the inverse rotation of the incident beam. The action of the amplitude matrices in the original and rotated reference frames ($\mathbf{A}$ and $\mathbf{A}'$, respectively) on the corresponding Jones vectors must lead to the same result, if the scattered direction is the same (its azimuthal scattering angle changes accordingly) [see Eq. (130)]. Thus, they are related as

$$\mathbf{A}' = e^{-in\psi} \mathbf{A} \mathbf{R}_{-\psi}, \tag{139}$$

i.e. the relation is different from that for the plane wave[51] only by a common phase factor, naturally related to the beam vorticity.

We further postulate that the Stokes vector is still defined by Eqs. (131)–(134), which implies the same relation between the amplitude and the scattering matrix, as for the plane wave. Thus, when the reference frame (or the scatterer) is rotated, the generalized Mueller matrix is transformed as

$$\mathbf{S}' = \mathbf{S} \begin{pmatrix} 1 & 0 & 0 & 0 \\ 0 & \cos(2\psi) & -\sin(2\psi) & 0 \\ 0 & \sin(2\psi) & \cos(2\psi) & 0 \\ 0 & 0 & 0 & 1 \end{pmatrix}. \tag{140}$$

As mentioned above, the main goal of the generalized Mueller calculus is to provide a quick solution for any linear combination of two basic beams. For instance, Eqs. (136) and (137) remain valid together with its meaning of scattering intensity in two planes for axisymmetric particles.

The main drawback of this generalization is that the Stokes vector of the vortex beam does not anymore have a clear physical meaning. The fundamental issue with measurability of this vector stems from the fact that the signal of the detector, illuminated by the Bessel beam, is not proportional to the detector area in contrast to the plane-wave case. In principle, one can normalize the reading of a large detector placed on the beam axis (integral of the Poynting vector) by inverse radius [cf. Eq. (114)]. According to Sections 3.4 and 3.5, such normalized intensity will be related to the Stokes parameter $I$. However, it is not clear how to adapt the standard techniques to measure other Stokes parameters (using linear and circular polarizers[23]) to vortex beams. Alternatively, one may relate the Stokes vector of the Bessel beam to that of a plane wave used to produce this beam, e.g., using an axicon. But we leave the analysis of both these approaches to measure generalized Stokes vector for future research. We also do not consider existing options to expand the Mueller calculus by increasing the size of the matrices, e.g. to separately account for transformation of each Fourier component of the incident field.[57]

Moreover, in the case of the TE and TM beams applying the above Mueller calculus will be misleading, since there are no two independent polarizations related by rotation [see Eq. (54)]. But as we showed previously [Eq. (55)], these beams are circular polarizations of the TEL or TML beams, respectively. And a similar correspondence holds for the Stokes vectors. Specifically,

$$\begin{pmatrix} E_{\parallel s} \\ E_{\perp s} \end{pmatrix}_{\text{TE,TM}} = E_0 \frac{e^{ik(r-z)}}{-ikr} \begin{pmatrix} S_2 & S_3 \\ S_4 & S_1 \end{pmatrix}_{\text{TEL,TML}} \begin{pmatrix} i \\ 1 \end{pmatrix}, \tag{141}$$

$$\begin{pmatrix} I_s \\ Q_s \\ U_s \\ V_s \end{pmatrix}_{\text{TE,TM}} = \frac{2|E_0|^2}{k^2 r^2} \begin{pmatrix} S_{11} & S_{12} & S_{13} & S_{14} \\ S_{21} & S_{22} & S_{23} & S_{24} \\ S_{31} & S_{32} & S_{33} & S_{34} \\ S_{41} & S_{42} & S_{43} & S_{44} \end{pmatrix}_{\text{TEL,TML}} \begin{pmatrix} 1 \\ 0 \\ 0 \\ -1 \end{pmatrix}, \tag{142}$$

where Eq. (141) is valid for the $yz$-scattering plane. Otherwise, additional phase factor is required according to Eqs. (138) or (54). However, this phase factor cancels out in the Mueller matrix, hence, Eq. (142) is valid for any scattering plane through the $z$-axis [cf. Eq. (140)].

Finally, the described generalization of the Mueller calculus if fully applicable to other vortex beams, as the formulas have only a single additional parameter – vorticity $n$, corresponding to the additional phase incurred by a full rotation over the beam axis.

### 4.2. Implementation in the ADDA code

Through the discussion in previous sections we showed that all known Bessel beams can be parametrized by their matrices **M** (Table 3). This parametrization constitutes the core of implementation of the Bessel beams in the ADDA code.[58] Any beam matrix can be specified through the command line:

```
-beam besselM <n> <α₀> <Re Me,x> <Re Me,y> <Re Mm,x> <Re Mm,y>
[<Im Me,x> <Im Me,y> <Im Mm,x> <Im Mm,y>]
```

where ⟨…⟩ denotes the values of the corresponding variables. Since matrix **M** is real for most of the Bessel beam types, only the real part of **M** is strictly required; the imaginary part is optional. Coordinates of the beam center relative to the particle center can be defined with a separate command line option (`-beam_center` …).

To simplify the code usage for standard Bessel beams, we also enabled separate specification of the LE, LM, CS, CS′, TEL, and TML types (although internally it is a wrapper substituting a specific matrix **M**). Those beams can be selected by the command line options

```
-beam bessel<X> <n> <α₀>,
```

where <X> is the beam identifier, for example "`-beam besselLE` …". We chose these beam types since they have well-defined parallel and perpendicular components (see Section 3.2). ADDA performs separate simulations for these two incident polarizations to calculate the amplitude and scattering matrices (see Section 4.1). The same happens for any other matrix **M** – it is assumed to define the perpendicular polarization, while the parallel one is taken from Eq. (53). In other words, matrix **M** (or beam type) is considered with respect to the beam reference frame, in which $yz$-plane is the scattering one. Then, in the case of a scatterer rotationally symmetric with respect to the beam axis, both amplitude and Mueller matrix are invariant with respect to the rotation of the scattering plane around this axis.

Technically speaking, ADDA performs 90° rotation of a perpendicular polarization with further correction of the phase to obtain the parallel one. Such transformation may lead to redundant calculations when **MR** is proportional to **M**, which is equivalent to **M** being a superposition of generalized circular polarizations of the same handiness [cf. Eq. (71)]. A specific case of the latter is the TE and TM beams, which can only be specified through their matrix **M**. However, an alternative option is to specify TEL or TML incident beam, respectively; the scattered Stokes vector can then be obtained from the computed Mueller matrix using Eq. (142). Finally, the initialization of the Bessel beam, i.e. computation of the incident electric field at each dipole position in ADDA, is negligibly fast. In most cases, it is faster than a single iteration of the iterative solver, used to determine the dipole polarizations.

## 5. Numerical results

First, we test our implementation of the Bessel beams in ADDA by comparing its results to reference data. The obvious choice for the reference method is the GLMT,[25] which allows one to calculate the scattering of sophisticated beams by particles with spherical symmetry. For instance, it was used to simulate the scattering of zero-order Bessel beam by a concentric sphere.[28] Hereinafter, we compare the scattering intensities in two planes: E-plane (parallel) and H-plane (perpendicular), given by Eqs. (136) and (137) respectively.

To quantify the discrepancy between the DDA and GLMT, we use the root mean square error (RMSE), defined as:[30]

$$\text{RMSE} = \frac{1}{\max|I_{\text{ref}}(\theta_i)|} \sqrt{\frac{1}{N_0} \sum_{i=1}^{N_0} (|I(\theta_i) - I_{\text{ref}}(\theta_i)|)^2}, \quad (143)$$

where $I$ and $I_{\text{ref}}$ are the scattering intensity calculated by the DDA and the GLMT, respectively, $N_0 = 181$ is the number of scattering angles.

The first test is on-axis scattering of the CS beam with the beam order $n = 0$ and half-cone angle $\alpha_0 = 15°$ by a concentric sphere with outer and inner radii equal to $\lambda$ and $\lambda/2$, respectively. The refractive indices of the shell and core are denoted as $m_1$ and $m_2$, respectively. We chose excessively fine discretization ($N_x = 128$ dipoles per the particle diameter) to avoid any concerns about the accuracy and used the default DDA formulation (so-called, lattice dispersion relation).[37] The results for the case $m_1 = m_2 = 1.33$, i.e., a homogeneous sphere, are shown in Figure 12. The small discrepancy is visible only in the backscattering, the overall RMSE is 1.6% for both scattering planes. The second case is $m_1 = 1.33$ and $m_2 = 1.55$ – the results are presented in Figure 13. Here the agreement is even better (within the size of symbols) with RMSE being 0.3% for both scattering planes.

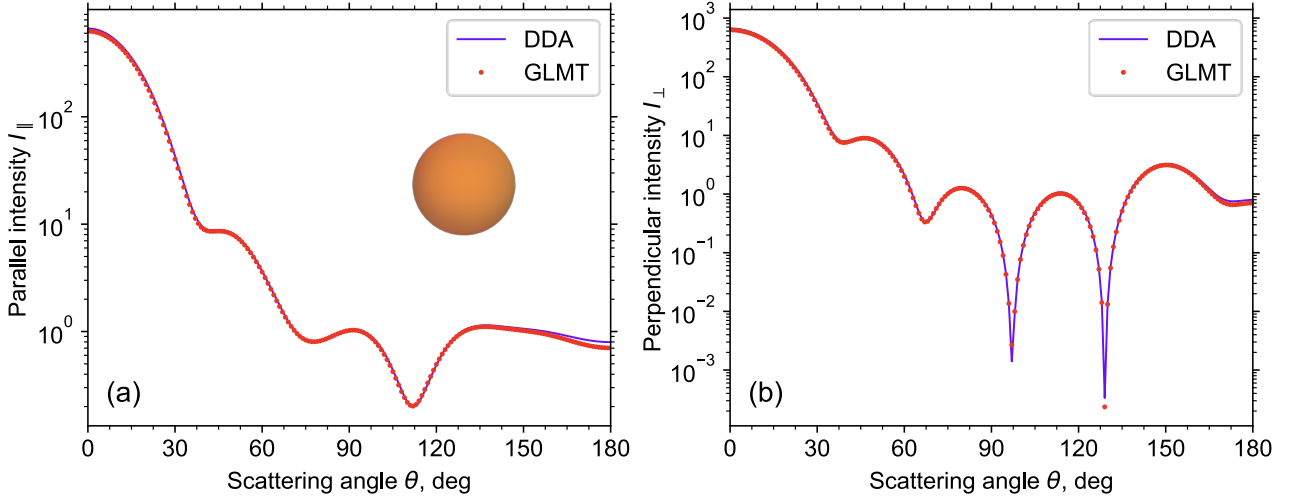

**Figure 12.** The comparison of the scattering intensities (a) $I_\parallel$ and (b) $I_\perp$ calculated with the DDA (our results) and GLMT (data from Ref. [28]) for scattering of the zero-order CS Bessel beam by a sphere. The logarithmic scale is used. See the text for details.

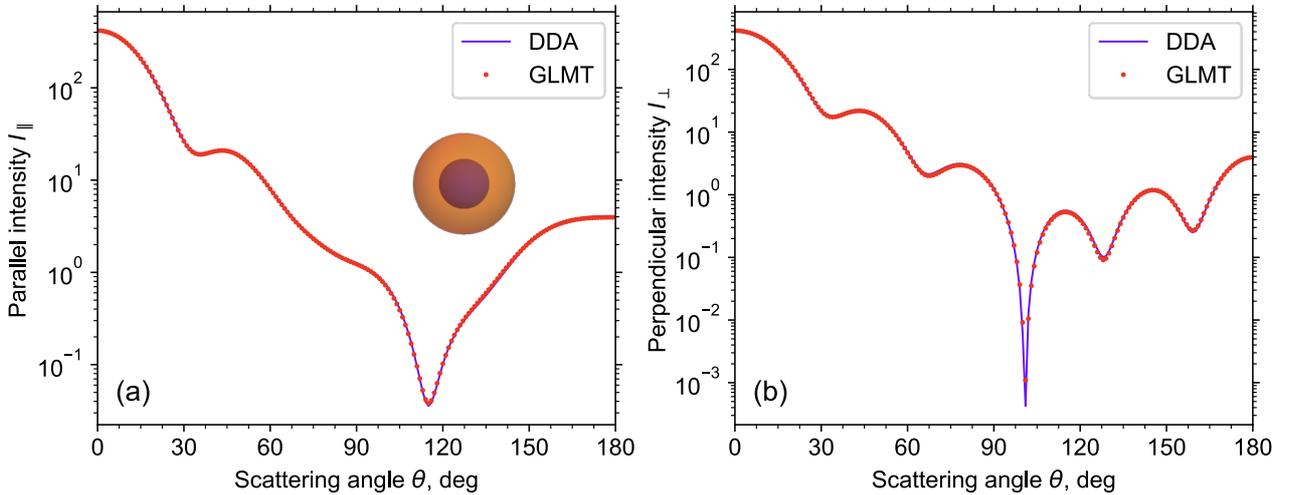

**Figure 13.** The same as Figure 12, but for a coated sphere. GLMT data are from Ref. [28].

While it is known that the DDA accuracy is commonly the worst at the backscattering direction,[59–61] we investigate the remaining discrepancy in Figure 12 in some details. For that we performed the DDA simulations for discretization from $N_x = 64$ to 256 and performed quadratic extrapolation of the backscattering intensity versus the discretization parameter $kdm_1 = 2\pi m_1/N_x$ (where $d$ is the dipole size) to the value of $d = 0$. The details of this procedure are described in.[59]

Note that the backscattering intensity is exactly the same for both scattering planes due to the rotational symmetry of the scatterer. The convergence curve and the extrapolation results (including 95% confidence interval) are shown in Figure 14 together with the reference value. One can see the DDA shows very smooth (and almost linear) convergence for the finer discretization, which is reflected by small uncertainty of the extrapolation result. to the reference value. Thus, about one third of the discrepancy between the DDA (for $N_x = 128$) and the GLMT at backscattering is due to the DDA error, while two thirds – due to the GLMT one. The latter can probably be decreases by tuning the parameters of that method.

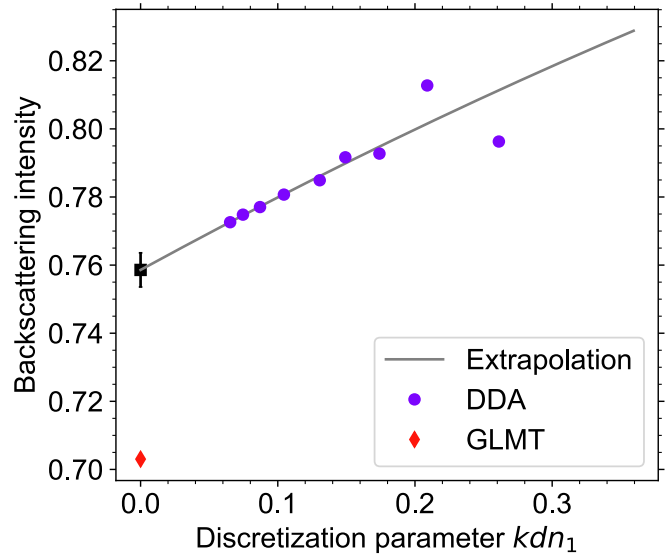

**Figure 14.** DDA convergence with refining discretization for backscattering intensity of the same sphere as in Figure 12. Quadratic extrapolation to $d = 0$ is performed and a 95% confidence interval for the corresponding value is also shown. See the text for details.

Second, to illustrate the DDA capabilities for non-spherical particles, we demonstrate the scattering of high-order Bessel beams of CS, CS′, TEL, and TML types ($n = 4$, $\alpha_0 = 45°$, and $\lambda = 632.8$nm) by a glass cube (cube side $a = 1$ μm, $m = 1.52$, $N_x = 15$) in Figure 15. The intensity profiles of these beams are also shown – they were calculated with ADDA itself using command line option `-store_beam`. This figure demonstrates the wide variation between the vector Bessel beam types of the same parameters. Naturally, the similarity in beam profiles leads to similarity in scattering intensities (e.g., TEL and TML types). Note also, that lack of global intensity maximum at the forward direction is due to the dark spot in the center of Bessel beams of non-zero order.

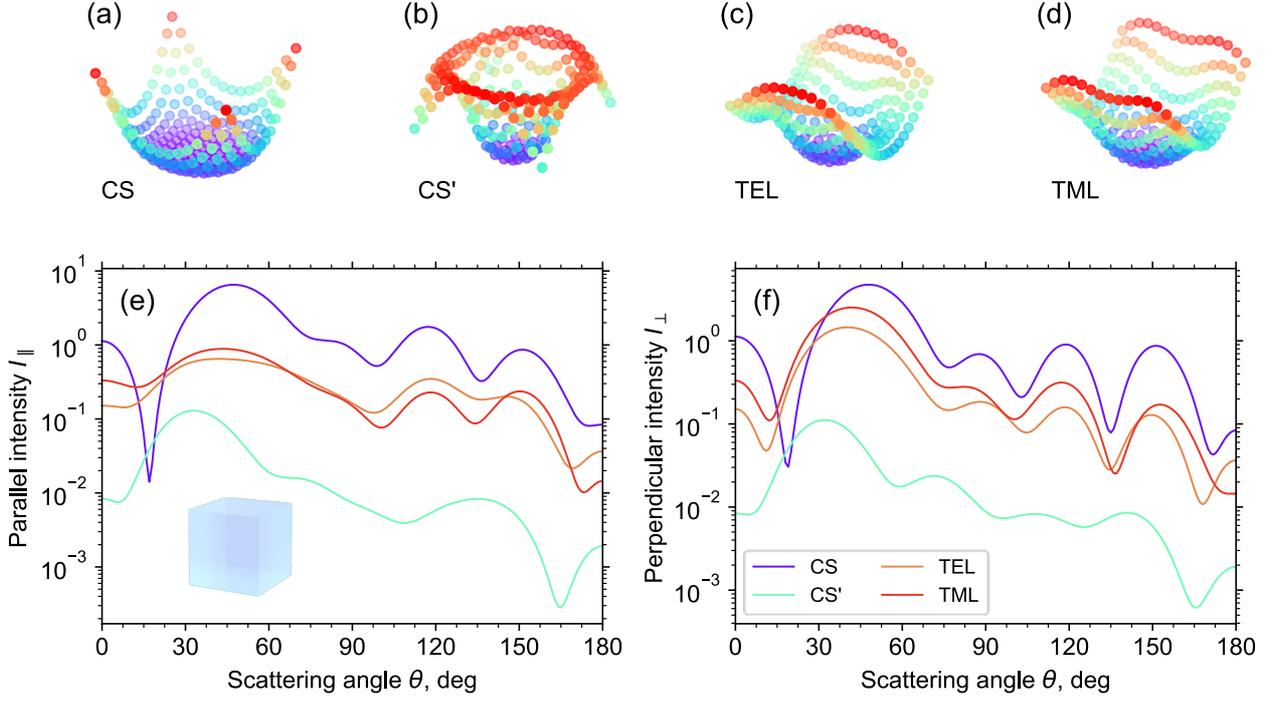

**Figure 15.** Intensity profiles for incident Bessel beam of CS (a), CS′ (b), TEL (c), and TML (d) types ($n = 4$, $\alpha_0 = 45°$, and $\lambda = 632.8\,\text{nm}$) over the central slice ($z = 0$) of the dipole grid for a cube ($a = 1\,\mu\text{m}, m = 1.52, N_x = 15$) calculated by ADDA. Also shown are the simulated scattering intensities (e) $I_\parallel$ and (f) $I_\perp$ for these scattering problems. See the text for details.

## 6. Conclusion

We have proposed a new approach to describe any vortex Bessel beam using a 2×2 matrix **M** comprised of transverse Hertz vector potentials. The duality and polarization rotations are then trivially expressed through the multiplication of **M** by rotation matrices from the left and right, respectively. Previously known beam types, specifically the TE, TM, LE, LM, and CS beams, are included as specific cases in this framework. And it additionally includes the new type with circular symmetric energy density, denoted as CS′. When combined with the CS type and two different polarizations, it leads to a complete basis of all Bessel beams, similar to the previously known bases of the TE, TM or LE, LM beams. For each beam type (except the TE and TM) we introduced the notation for elliptical polarization ($\alpha, \beta$). This definition also applies to the newly introduced linearly polarized components of the TE and TM beams, denoted as TEL and TML. The latter two also form a complete basis for Bessel beams, while their circular polarizations result in the TE and TM beams, respectively, of adjacent orders. The generalized elliptical polarizations are tightly connected with the beam rotations, which are expressed by standard rotation matrices with additional phase factor depending on the beam order $n$ (vorticity). The orthogonal polarizations for linearly polarized beam types are then defined accounting for this phase factor in agreement with the existing $x$- and $y$-polarizations of some beam types in the literature.

To gain more insight into both previously known and new Bessel beam types, we described their behavior in two limiting cases: the plane-wave ($n = 0$, $\alpha_0 \to 0°$) and bullet limits ($\alpha_0 \to 90°$). As expected, some of the beam types vanish in the plane-wave limit, while others become equivalent to each other. We also derived expressions for quadratic functionals of the fields (such as the energy density and Poynting vector) for a beam with arbitrary matrix **M** and derived orthogonality relations between various beam types. For the latter we introduced a normalization of the integrals, since the ideal Bessel beams are not square-integrable. The orthogonality relations complete the discussion of various bases for the Bessel beams. In particular, the basis of LE, LM beams, which directly corresponds to the elements of matrix **M**, is not completely orthogonal. Thus, we constructed a general unified framework for description of Bessel beams and provided a comprehensive reference of all their properties that may be relevant for any applications.

To enable simulation of scattering of Bessel beams by arbitrary particles, we, first, generalized the classical Mueller calculus to this case. We defined the amplitude matrix in terms of two orthogonal polarizations for the same beam type (as discussed above), which leads to the simplest transformation rule under the rotation of the scattering plane around the beam axis. This rule differs from the case of a plane wave scattering only by a constant phase factor that cancels out in the rotation transformations of Mueller matrices. Importantly, exactly the same generalized definitions can be used for any vortex beams.

Finally, we have modified the open-source code ADDA to enable simulation of Bessel-beam scattering. Apart from the beam order, half-cone angle, and its position relative to the particle, a user need to specify either the beam type (one of the above options) or a complete matrix **M**. ADDA then computes the angle-resolved Mueller matrix, which can be used to compute the scattering of any elliptical polarization of the same beam without any extra efforts. The simulated scattering intensity of zero-order Bessel beam by a homogeneous or coated spheres shows a perfect agreement with the reference data, obtained with the GLMT. Thus, the new version of ADDA allows easy and efficient simulation of scattering of any Bessel beam by a particle with arbitrary shape and internal structure. This can be used to supplement theoretical analysis of Bessel beams, including their symmetry, as well as for advancing various practical applications of such beams. Promising future research directions include consideration of Bessel-beam scattering near a plane substrate and efficient calculation of optical forces (e.g., in optical tweezers).

## 7. Acknowledgments

We thank Gérard Gouesbet, James A. Lock, and Vadim A. Markel for insightful discussions and Zhuyang Chen for providing the raw GLMT data plotted in Figure 12 and Figure 13. We are also grateful to Alexander Kichigin for providing a preliminary version of the script used for extrapolation and his help in implementing Bessel beams in ADDA. This work is supported by the Russian Science Foundation (Grant No. 18-12-00052).

## 8. Associated content

Figures 12–15 and the underlying DDA data can be reproduced by Python scripts in the ADDA repository – https://github.com/adda-team/adda/tree/master/examples/papers/2022_bessel.